%% file: main.tex
\documentclass[a4paper,11pt,twoside]{article}
\usepackage[paperheight=29.7cm,paperwidth=21cm,outer=2.25cm,inner=2.25cm,top=2cm,bottom=2cm,headheight=13.6pt,includehead,includefoot]{geometry} 
\usepackage{silence}
\newcommand{\textlineskip}{\baselineskip=13pt}

\providecommand{\keywords}[1]{\noindent\textbf{Keywords:} #1}

\def\fnm#1{$^{\mbox{\scriptsize #1}}$}
\def\fnt#1#2{\footnotetext{\kern-.3em%
          {$^{\mbox{\scriptsize #1}}$}{#2}}}
          
\newcommand{\fcaption}[1]{\caption{#1}}
\newcommand{\tcaption}[1]{\caption{#1}}

\usepackage{fancyhdr}

\fancypagestyle{pagestyle}{%
  \fancyhf{}
  \fancyhead[OR]{E. Zardini, M. Rizzoli, S. Dissegna, E. Blanzieri, D. Pastorello}
  \fancyhead[EL]{Reconstructing Bayesian Networks on a Quantum Annealer}
  \fancyfoot[C]{\thepage}%
}

\fancypagestyle{firstpagestyle}{%
  \fancyhf{}
  
  \fancyfoot[C]{\thepage}
}

\renewcommand{\thefootnote}{\fnsymbol{footnote}}  

\usepackage{amsmath}
\usepackage{amssymb}
\usepackage[linesnumbered]{algorithm2e}
\usepackage[font={footnotesize}]{caption}

\usepackage{bm}
\input{symbols_and_commands}

\usepackage{multirow}
\usepackage[figuresright]{rotating}

\usepackage{url}
\usepackage[hyperfootnotes=false]{hyperref}
\usepackage{xcolor}
\hypersetup{
  colorlinks,
  linkcolor={black},
  citecolor={black},
  urlcolor={black}
}
\usepackage[noabbrev,capitalise]{cleveref}

\usepackage{soul}

\usepackage{etoolbox}
\usepackage{accents}
\robustify{\underaccent}

\begin{document}

\normalsize\textlineskip
\thispagestyle{firstpagestyle}
\setcounter{page}{1}

\centerline{\Large\bf
Reconstructing Bayesian Networks on a}
\vspace*{0.035truein}
\centerline{\Large\bf Quantum Annealer}

\vspace{25pt}

\begin{center}
    \textbf{Enrico Zardini$^{\star \dagger}$}, \hspace{1pt}
    \textbf{Massimo Rizzoli$^\star$}, \hspace{1pt}
    \textbf{Sebastiano Dissegna$^\star$}, \\  
    \textbf{Enrico Blanzieri$^{\star \ddagger}$}, \hspace{1pt}
    \textbf{Davide Pastorello$^{\star \ddagger}$}

    \vspace*{10pt}

    $^\star$ Department of Information Engineering and Computer Science\\ University of Trento \\ 
    $ $ via Sommarive 9, 38123 Povo, Trento, Italy

    \vspace*{10pt}

    $^\dagger$ enrico.zardini@unitn.it

    \vspace*{10pt}

    $^\ddagger$ Trento Institute for Fundamental Physics and Applications \\ 
    $ $ via Sommarive 14, 38123 Povo, Trento, Italy
\end{center}

\vspace*{15pt}

\begin{abstract}
\noindent
Bayesian networks are widely used probabilistic graphical models, whose structure is hard to learn starting from the generated data. O'Gorman et al. have proposed an algorithm to encode this task, i.e., the Bayesian network structure learning (BSNL), into a form that can be solved through quantum annealing, but they have not provided an experimental evaluation of it. In this paper, we present (i) an implementation in Python of O'Gorman's algorithm, (ii) a divide et impera approach that allows addressing BNSL problems of larger sizes in order to overcome the limitations imposed by the current architectures, and (iii) their empirical evaluation. Specifically, several problems with an increasing number of variables have been used in the experiments. The results have shown the effectiveness of O'Gorman's formulation for BNSL instances of small sizes, and the superiority of the divide et impera approach on the direct execution of O'Gorman's algorithm.

\vspace*{10pt}
\end{abstract}

\setcounter{footnote}{0}
\renewcommand{\thefootnote}{\alph{footnote}}

\vspace*{1pt}\textlineskip    

\section{Introduction}
\label{sec:intro}
\noindent
Bayesian networks (BNs) are graphical probabilistic models in which the joint density distribution of multiple random variables is represented over a directed acyclic graph \cite{pearl1985bayesian}. In detail, each variable corresponds to a node of the graph, and the overall joint density distribution is obtained by multiplying the conditional density distribution of each variable given its parents on the graph. As a consequence, the topology of the graph defines the independence conditions, i.e., a variable is independent of its non-descendants given its parents. BNs are widely used for representing uncertain domains and their structure allows for probabilistic reasoning. Obtaining a BN representation from data is a learning task with a long history; in particular, the subtask of learning the topology, also known as BN reconstruction, has received much attention \cite{spirtes1995learning}, especially when BNs are used to represent causal relationships \cite{pearl1995bayesian}. Moreover, some recent papers have dealt with the possible application of quantum computing to Bayesian networks \cite{ogorman, ozols, borujeni}.

Quantum computing (QC) is a kind of computation that exploits quantum mechanical phenomena for information processing, and, nowadays, working quantum computers are available on the market \cite{dwave}. QC will have an impact on artificial intelligence and machine learning. Indeed, it has the potentiality to allow efficient solutions to many of the search and optimization problems encountered in these fields. In the last years, some applications of quantum computing to Bayesian networks have been proposed, such as the following: a method for learning the structure of a BN using a quantum annealer \cite{ogorman}; an algorithm for Bayesian inference based on amplitude amplification, which is a quantum version of the classical rejection sampling algorithm used for inference in Bayesian networks \cite{ozols}; a systematic method for designing a quantum circuit to represent a generic discrete BN \cite{borujeni}. In particular, the proposal of O’Gorman et al. \cite{ogorman} considers a quantum annealing architecture instead of a gate-based quantum computer. Quantum annealing is a type of heuristic search for solving optimization problems by finding the low-energy states of a quantum system \cite{kadowaki}, and quantum annealers are non-universal specific-purpose quantum computers implementing quantum annealing. The advantage of the existing quantum annealers lies in the high number of qubits w.r.t. the available prototypes of general-purpose quantum computers. The paper by O’Gorman et al. describes an effective encoding of the BN reconstruction problem into a quantum annealer architecture. However, no implementation and empirical evaluation on a real quantum machine are provided.

In this paper, we present an empirical evaluation of the proposal of O'Gorman et al. in order to assess its practical applicability using the available architectures. Since the problem encoding and the subsequent embedding in the quantum architecture limit the direct application to around 18 Bayesian variables (at time of writing), we also propose a divide-et-impera approach to overcome this limitation. Both the original algorithm and the new scheme have been tested on different problems with a growing number of variables. The code is available under the GPLv2 licence \cite{ogorman_implementation_github,divide_et_impera_implementation_github}.

The paper is organized as follows: \cref{sec:background} provides some background information; \cref{sec:ogorman-implementation} describes the implementation of O'Gorman's algorithm \cite{ogorman}; \cref{sec:divide-et-impera} presents the so-called \textit{divide et impera} approach; \cref{sec:empirical-evaluation} is devoted to the empirical evaluation; \cref{sec:conclusion} contains the concluding remarks.

\section{Background}
\label{sec:background}
\noindent
This section provides information about QUBO problems, quantum annealing and D-Wave, the embedding into quantum processing units (QPUs), the Bayesian network structure learning problem, and O'Gorman's QUBO algorithm \cite{ogorman} to address it.

\subsection{QUBO problems}
\label{subsec:qubo-problems}
\noindent
Quadratic Unconstrained Binary Optimization (QUBO) problems are optimization problems of the form
\begin{equation}
    \label{eq:xtQx}
    \arg\displaystyle\min_{x} x^T Q x
\end{equation}
where $x$ is a binary vector, and $Q$ is an upper triangular (or symmetric) matrix of real values. In particular, let $x$ be an $n\times1$ vector and $Q$ a $n\times n$ upper triangular matrix; then, it is possible to rewrite the QUBO problem as follows:
\begin{align}
  x^T Q x
          &= \displaystyle\sum_{i=1}^n q_{ii}x_{i}^2 + \displaystyle\sum_{i=1}^n\displaystyle\sum_{j=i+1}^n q_{ij} x_i x_j \nonumber \\
          &= \displaystyle\sum_{i=1}^n q_{ii}x_{i} + \displaystyle\sum_{i=1}^n\displaystyle\sum_{j=i+1}^n q_{ij} x_i x_j
\end{align}
where $x_i^2 = x_i$ since $x_i \in \mathbb{B}= \{0, 1\}$. In practice, the main diagonal of $Q$ contains the linear coefficients ($q_{ii}$), whereas the rest of the matrix contains the quadratic ones ($q_{ij}$). Although QUBO problems are unconstrained by definition, it is actually possible to introduce constraints by representing them as penalties. Several examples are provided by Glover et al.~\cite{glover}.

The significance of the QUBO formulation mainly lies in its computational equivalence with the Ising model, which is the physical model upon which annealers are built. The only difference is the domain of variables: $\{0, 1\}$ for the QUBO formulation, $\{-1, +1\}$ for the Ising one. Hence, by applying a trivial conversion, it is possible to exploit quantum annealers to solve problems expressed as QUBO.

\subsection{Quantum annealing and D-Wave machine}
\label{subsec:quantum-annealing-and-dwave}
Quantum annealing (QA) is a heuristic search used to solve optimization problems~\cite{kadowaki}. The solution of a given problem corresponds to the \emph{ground state} (the less energetic physical state) of a $n$-qubit system with energy described by a \emph{problem Hamiltonian} $H_P$, which is a Hermitian $2^n\times 2^n$ matrix.
The annealing procedure is implemented by a time evolution of the quantum system towards the ground state of the problem Hamiltonian. More precisely, let us consider the time-dependent Hamiltonian
\begin{equation}
H(t)=\Gamma(t) H_D+H_P,
\end{equation}
where $H_P$ is the problem Hamiltonian, and $H_D$ is the \emph{transverse field Hamiltonian}, which gives the kinetic term inducing the exploration of the solution landscape by means of quantum fluctuations. $\Gamma$ is a decreasing function that attenuates the kinetic term, driving the system towards the global minimum of the problem landscape represented by $H_P$.

QA can be physically realized by considering a quantum spin glass, which is a network of qubits arranged on the vertices of a graph $\langle V,E\rangle$, with $|V|=n$ and whose edges $E$ represent the couplings among the qubits. The problem Hamiltonian is defined as
\begin{equation}\label{HP}
H_P=H(\boldsymbol\Theta)=\sum_{i\in V} \theta _i \sigma_z^{(i)} +\sum_{(i,j)\in E} \theta_{ij}\sigma_z^{(i)} \sigma_z^{(j)},
\end{equation}
where the real coefficients $\theta_i, \theta_{ij}$ are arranged into the matrix $\boldsymbol\Theta$. $H(\boldsymbol\Theta)$ is an operator on the $n$-qubit Hilbert space $\mathsf H=(\mathbb C^2)^{\otimes n}$, whereas $\sigma_z^{(i)}$ acts as the Pauli matrix
\begin{equation}
\sigma_z=\left(\begin{matrix}
1 & 0\\
0 & -1
\end{matrix}
\right)
\end{equation}
on the $i$th tensor factor and as the $2\times 2$ identity matrix on the other tensor factors.
Regarding the coefficient matrix $\boldsymbol\Theta$, it is the $n\times n$ symmetric square matrix of real coefficients of $\mathsf E$ (called \emph{weights}) defined as
\begin{equation}\label{w}
\boldsymbol\Theta_{ij}:=\left\{
\begin{array}{ll}
\theta_i, & i=j,\\
\theta_{ij}, & (i,j)\in E,\\
0, & (i,j)\not\in E,
\end{array}\right.
\end{equation}
with $\theta_i$ physically corresponding to the local field on the $i$th qubit, and $\theta_{ij}$ to the coupling between the qubits $i$ and $j$.
In particular, the Pauli matrix $\sigma_z$ has two eigenvalues $\{-1,1\}$, which correspond to the binary states, \emph{spin down} and \emph{spin up}, of each qubit. Thus, the spectrum of eigenvalues of the problem Hamiltonian (Eq.~\ref{HP}) is the set of all possible values of the cost function given by the energy of the well-known \emph{Ising model}:
\begin{equation}\label{costf}
    \mathsf E(\boldsymbol{\Theta}, \boldsymbol z)=\sum_{i\in V} \theta_i z_i +\sum_{(i,j)\in E} \theta_{ij}z_i z_j,\quad \boldsymbol z=(z_1,...,z_n)\in\{-1,1\}^{|V|}.
\end{equation}
In practice, the annealing procedure, also called \emph{cooling}, drives the system into the ground state of $H(\Theta)$, which corresponds to the spin configuration encoding the solution:
\begin{equation}\label{argmin_zE}
\boldsymbol z^*=\arg\!\!\!\!\!\!\!\!\min_{\boldsymbol z\in\{-1,1\}^{|V|}} \mathsf E(\boldsymbol\Theta,\boldsymbol z).
\end{equation}
Given a problem, the annealer is initialized using a suitable choice of the weights $\boldsymbol\Theta$, and the binary variables $z_i\in\{-1,1\}$ are physically realized by the outcomes of the measurements performed on the qubits located in the vertices $V$. In order to solve a general optimization problem through QA, it is first necessary to find an \textit{encoding} of the objective function in terms of the cost function~\eqref{costf}, which is not easy in general.

D-Wave Systems is a Canadian company producing quantum annealers, i.e., physical machines implementing the quantum annealing process. Currently, the available models are the D-Wave 2000Q, exploiting the \textit{Chimera} topology, and the D-Wave Advantage, featuring the \textit{Pegasus} topology. The former has 2048 qubits, each connected to 6 other qubits, whereas the latter has 5640 qubits, each connected to 15 other qubits. A higher amount of qubits allows for larger problems to be submitted, but the most relevant feature is the connectivity, which determines the complexity of the representable problems. For these reasons, the D-Wave Advantage has been chosen for the experiments.

\subsection{Quantum processing unit (QPU) embedding}
\label{subsec:qpu-embedding}
\noindent
To practically use quantum annealing for solving QUBO problems, the problem variables must be mapped to the QPU qubits. However, due to the sparseness of the available annealer topologies, a direct representation of the problem is typically not possible. The solution consists in chaining together multiple physical qubits that will act as a single logical qubit. In this way, the connectivity of the annealer graph is increased at the price of reducing the number of logical qubits available and, consequently, the size of representable problems. The entire process is known as embedding or \textit{minor embedding} (in the glossary of D-Wave) \cite{minor_embedding, minor_embedding_example}. In particular, D-Wave's Ocean library provides the \textit{EmbeddingComposite} class \cite{embedding_composite} to automatically perform the minor embedding of the supplied QUBO matrices, and a new embedding is computed for every annealer read.

\subsection{Bayesian network structure learning (BSNL)}
\label{subsec:bnsl}
\noindent
A Bayesian network (BN) is a directed acyclic graph (DAG) representing the conditional dependencies of a set of random variables. In particular, the nodes of the graph represent the variables, whereas the edges represent the conditional dependencies between them. Moreover, each node is associated with the conditional probability distribution of the node itself given its parents.

The method proposed by O'Gorman et al. \cite{ogorman} focuses on the network structure learning (BNSL) problem, which consists in finding the Bayesian network that most likely has generated a given set of data. The problem is NP-Complete \cite{bnsl_np_complete}, and the authors expect a polynomial speedup using quantum annealing. In detail, to take advantage of the new technology, a hardware compatible QUBO formulation of the BNSL problem is provided in the paper together with sufficient lower bounds for the penalties.

More formally, a Bayesian network can be defined as a pair $(B_s,B_p)$, where $B_s$ is a DAG and $B_p$ is the set of associated conditional probabilities. Then, given a database $D=\{\mathbf{x}_i|1 \leq i \leq N\}$ with $\mathbf{x}_i$ representing the state of all variables, the objective consists in finding the structure that maximises the posterior probability distribution $p(B_s|D)$. However, due to the proportionality of $p(B_s|D)$ and $p(D|B_s)$ by Bayes' Theorem, it is possible to reformulate the problem as maximizing $p(D|B_s)$, which is given by
\begin{equation}
  \label{eq:p-d-given-b}
  p(D|B_s) = \displaystyle\prod_{i=1}^n \displaystyle\prod_{j=1}^{q_i} \frac{\Gamma(\alpha_{ij})}{\Gamma(N_{ij} + \alpha_{ij})} \displaystyle\prod_{k=1}^{r_i} \frac{\Gamma(N_{ijk} + \alpha_{ijk})}{\Gamma(\alpha_{ijk})},
\end{equation}
where $\Gamma$ is the gamma function, $q_i$ is the number of joint states of the parent set of the $i$-th random variable, $r_i$ is the number of states of the random variable itself, $N_{ijk}$ is the number of occurrences in $D$ with the $i$-th random variable in its $k$-th state and the variable's parent set in its $j$-th state, $\alpha_{ijk}$ is the hyperparameter of the assumed Dirichlet prior for the node's conditional probability distribution, $N_{ij}$ and $\alpha_{ij}$ are the sums of the corresponding parameter values over $k$.

\subsubsection{QUBO formulation of BNSL}
\label{subsubsec:qubo-formulation-of-bnsl}
In their work \cite{ogorman}, O'Gorman et al. provide a Hamiltonian function for the BNSL problem. Given the BNSL Hamiltonian, the construction of the QUBO matrix is straightforward: it is sufficient to map the coefficients of the variables into the matrix entries. In particular, the BNSL Hamiltonian consists of three components: the score Hamiltonian ($H_{score}$), which is responsible for evaluating the quality of the solution graph; the max Hamiltonian ($H_{max}$), which is in charge of penalising the solutions including nodes with a number of parents greater than $m$, a constraint dictated by resource limits; the cycle Hamiltonian ($H_{cycle}$), further divided in consistency Hamiltonian ($H_{consist}$) and transitivity Hamiltonian ($H_{trans}$), which penalises the solutions containing cycles. Hence, the full Hamiltonian ($H$) is given by
\begin{equation}
   \label{eq:full-bsnl-hamiltonian}
   H(\mathbf{d},\mathbf{y},\mathbf{r}) = H_{score}(\mathbf{d}) + H_{max}(\mathbf{d},\mathbf{y}) + H_{cycle}(\mathbf{d},\mathbf{r}),
\end{equation}
where $\mathbf{d}$ corresponds to the $n(n-1)$ bits used to represent the presence/absence of edges between nodes, whereas $\mathbf{y}$ and $\mathbf{r}$ are additional variables exploited to encode the constraints.

\paragraph{Score Hamiltonian}
The score Hamiltonian ($H_{score}$) is calculated separately for each variable, and the components are then summed together. In detail, the score Hamiltonian for the $i$-th variable is given by
\begin{equation}
  \label{eq:score-hamiltonian-formula}
  H_{score}^{(i)}(\mathbf{d}_i) = \displaystyle\sum_{\underaccent{\lvert J \rvert \leq m}{J \subset \{1..n\} \setminus \{i\}}}\left(w_i(J) \displaystyle\prod_{j \in J}d_{ji}\right),
\end{equation}
where $\mathbf{d}_i$ includes all the bits ($d_{ji}$) encoding edges towards the considered node, $m$ is the largest allowed size for the parent set, and $w_i$ is computed as follows:
\begin{equation}
  \label{eq:w-formula}
  w_i(J) = \displaystyle\sum_{l=0}^{\lvert J \rvert}(-1)^{\lvert J \rvert - l}\displaystyle\sum_{\underaccent{\lvert K \rvert = l}{K \subset J}}s_i(K),
\end{equation}
with $s_i$ being score values obtained from Eq.~\eqref{eq:p-d-given-b}, introducing a logarithm for numerical efficiency. Specifically, $s_i$ is given by
\begin{equation}
  \label{eq:s-formula}
  s_i(\varPi_i(B_s)) = - \log\left(\displaystyle\prod_{j=1}^{q_i}\frac{\Gamma(\alpha_{ij})}{\Gamma(N_{ij} + \alpha_{ij})} \displaystyle\prod_{k=1}^{r_i} \frac{\Gamma(N_{ijk} + \alpha_{ijk})}{\Gamma(\alpha_{ijk})}\right),
\end{equation}
where $\varPi_i(B_s)$ denotes the parent set of the $i$-th node. In practice, the sum of the $s_i$ values is equal to $- \log p(D|B_s)$.

\paragraph{Max Hamiltonian}
Analogously to the score Hamiltonian, the max Hamiltonian is computed separately for each variable as
\begin{equation}
  \label{eq:max-hamiltonian-formula}
  H_{max}^{(i)}(\mathbf{d}_i,\mathbf{y}_i) = \delta_{max}^{(i)}(m - d_i - y_i)^2,
\end{equation}
where $\delta_{max}^{(i)} > 0$ is the penalty weight, $d_i$ is the in-degree of the considered node (given by $\sum_{1 \leq j \leq n\,\cap\,j\neq i}d_{ji}$), and $y_i \in \mathbb{Z}$ is a slack variable (encoded via binary expansion in $\mathbf{y}_i$ using $\mu$ bits~\fnm{a}\fnt{a}{ $y_i = \sum_{l=1}^{\mu}2^{l-1}y_{il}$, with $\mu = \lceil \log_{2}(m+1) \rceil$}) that allows $H_{max}^{(i)}$ being zero when the constraint is satisfied. Indeed, $H_{max}^{(i)}$ is zero if the considered node has at most $m$ parents, otherwise it carries a positive penalty.

\paragraph{Cycle Hamiltonian}
As mentioned previously, the cycle Hamiltonian is defined as the sum of two components:
\begin{equation}
  \label{eq:cycle-hamiltonian-formula}
  H_{cycle}(\mathbf{d},\mathbf{r}) = H_{trans}(\mathbf{r}) + H_{consist}(\mathbf{d},\mathbf{r}),
\end{equation}
where $\mathbf{r}$ represents $n(n-1)/2$ additional boolean variables encoding a topological order ($r_{ij}$ is $1$ if the $i$-th node precedes the $j$-th one, $0$ otherwise). In detail, the transitivity Hamiltonian penalises the cycles of length three in the $r_{ij}$ values, and is computed separately for each possible 3-set of variables as
\begin{equation}
  \label{eq:trans-hamiltonian-formula}
  H_{trans}^{(ijk)}(r_{ij}, r_{jk}, r_{ik}) = \delta_{trans}^{(ijk)} \left(r_{ik} + r_{ij}r_{jk} - r_{ij}r_{ik} - r_{jk}r_{ik}\right),
\end{equation}
where $\delta_{trans}^{(ijk)}$ is the positive penalty added if the $i$-th, $j$-th and $k$-th variables form a 3-cycle. As in the previous cases, the $H_{trans}^{(ijk)}$ components are summed up to obtain the full $H_{trans}$. Instead, the consistency Hamiltonian penalises the solutions for which the topological order contained in $\mathbf{r}$ is inconsistent with the graph structure encoded in $\mathbf{d}$. In practice, it makes disadvantageous the solutions in which $r_{ij} = 1$ and $d_{ji} = 1$, or $r_{ij} = 0$ and $d_{ij} = 1$. The Hamiltonian is computed separately for each pair of variables as
\begin{equation}
  \label{eq:consist-hamiltonian-formula}
  H_{consist}^{(ij)}(d_{ij},d_{ji},r_{ij}) = \delta_{consist}^{(ij)} (d_{ji}r_{ij} + d_{ij} - d_{ij}r_{ij}),
\end{equation}
where $\delta_{consist}^{(ij)}$ is the positive penalty associated with the inconsistency. It is also worth highlighting that the penalties $\delta_{trans}^{(ijk)}$ and $\delta_{consist}^{(ij)}$ are invariant to the permutation of the superscript indices (the set of variables remains the same).

\subsubsection{QUBO size and penalty values}
\label{subsubsec:qubo-size-n-penalty-vals}
The QUBO formulation of the BNSL problem consist of $n(n-1)$ binary variables ($d_{ij}$) encoding the graph structure, $n\mu = n\lceil \log_{2}(m+1) \rceil$ binary slack variables ($y_{il}$) related to the maximum parent constraint, and $n(n-1)/2$ binary variables ($r_{ij}$) encoding a topological order (related to the absence of cycles constraint). Hence, the QUBO encoding of $n$ Bayesian variables requires $\mathcal{O}(n^2)$ binary variables. Nevertheless, since $H_{score}$ contains multiplications with $m$ factors, if $m \geq 3$, additional steps and slack variables are needed to convert the problem into a quadratic equation. For instance, according to O'Gorman et al., $n \lfloor \frac{(n-2)^2}{4} \rfloor$ binary slack variables are needed to reduce a BNSL problem with $m = 3$ to a quadratic form, increasing the total number of binary variables to $\mathcal{O}(n^3)$. In this work, only the formulation for $m=2$ has been used in the experiments, although some problems taken into account have more than two parents per node.

Concerning the penalty values, O'Gorman et al. provide the following lower bounds (they have also demonstrated their sufficiency):
\begin{equation}
  \label{eq:delta-max-formula}
  \delta_{max}^{(i)} > \displaystyle\max_{j \neq i}\Delta_{ji},\,1 \leq i \leq n,
\end{equation}
\begin{equation}
  \label{eq:delta-consist-formula}
  \delta_{consist}^{(ij)} > (n-2)\displaystyle\max_{k \notin \{i,j\}}\delta_{trans}^{(ijk)},\,1 \leq i < j\leq n,
\end{equation}
\begin{equation}
  \label{eq:delta-trans-formula}
  \delta_{trans}^{(ijk)} = \delta_{trans} > \displaystyle\max_{\underaccent{i'\neq j'}{1 \leq i',j' \leq n}}\Delta_{i'j'},\,1 \leq i < j < k \leq n,
\end{equation}
where $\Delta_{ji}$ is an estimate of the largest increase in score due to the insertion of an arc from the $j$-th to the $i$-th node. In the case of $m = 2$, it is given by
\begin{equation}
  \label{eq:delta-formula}
  \Delta_{ji} = \max\{0, \Delta'_{ji}\}, \\
\end{equation}
\begin{equation}
  \label{eq:delta-prime-formula}
  \Delta'_{ji} = -w_i(\{j\}) - \displaystyle\sum_{\underaccent{k \neq i,j}{1 \leq k \leq n}}\min\{0,w_i(\{j,k\})\}.
\end{equation}
Instead, for $m \geq 3$, computing $\Delta_{ji}$ becomes an intractable optimization problem.

\section{O'Gorman's Algorithm Implementation}
\label{sec:ogorman-implementation}
\noindent
A Python implementation of O'Gorman's algorithm, which provides a way to build the QUBO matrix for a BNSL problem, has been developed in this work. Since the D-Wave's Ocean suite, which is necessary for interacting with the quantum annealer, is implemented in Python, the programming language in question has turned out to be the most reasonable choice. This section provides the implementation details, some considerations about the complexity of the implementation, and the description of a method to speed up the execution.

\begin{algorithm}[htb]
  \footnotesize

  \KwIn{number of Bayesian variables $n$, list $r = (r_i)_{i=1}^n$ with $r_i$ being the number of states of the $i^{th}$ variable, dataset $examples$}
  \KwResult{QUBO matrix $Q$}

  \tcp{calculation of the values needed to construct $Q$}
  $parentSets \gets calcParentSets(n)$\;
  $\alpha \gets calcAlpha(n,r,parentSets)$\;
  $s \gets calcS(n,r,parentSets,\alpha,examples)$\tcp*{Algorithm \ref{alg:calcS}}
  $w \gets calcW(n,parentSets,s)$\tcp*{Algorithm \ref{alg:calcW}}
  $\Delta \gets calcDelta(n,parentSets,w)$\tcp*{Eq.~\eqref{eq:delta-formula} and \eqref{eq:delta-prime-formula}}
  $\delta_{max} \gets calcDeltaMax(n,\Delta)$\tcp*{Eq.~\eqref{eq:delta-max-formula}}
  $\delta_{trans} \gets calcDeltaTrans(n,\Delta)$\tcp*{Eq.~\eqref{eq:delta-trans-formula}}
  $\delta_{consist} \gets calcDeltaConsist(n,\delta_{trans})$\tcp*{Eq.~\eqref{eq:delta-consist-simplified-calc}}
  \tcp{construction of $Q$}
  $Q \gets zeroMatrix()$\;
  $Q \gets fillQ(Q,n,parentSets,w,\delta_{max},\delta_{trans},\delta_{consist})$\tcp*{Algorithm \ref{alg:fillQ}}

  \Return{Q}\;

  \caption{\it calcQUBOMatrix(n, r, examples)}
  \label{alg:calcQUBOMatrix}
\end{algorithm}

\subsection{QUBO matrix construction}
\label{subsec:qubo-matrix-construction}
The pseudocode of the implementation of O'Gorman's algorithm is shown in \cref{alg:calcQUBOMatrix}, which includes calls to Algorithms \ref{alg:calcS}-\ref{alg:calcW}-\ref{alg:fillQ}. In particular, the main algorithm takes as input the number of Bayesian variables $n$, the number of states $r_i$ for each variable, and the dataset of examples, and produces as output the QUBO matrix $Q$ that represents the considered BNSL problem.

Before effectively building the matrix, it is necessary to compute several intermediate values. First, all possible parent sets ($\varPi_i(B_s)$ in O'Gorman's formulation) are calculated for each Bayesian variable. The maximum number of parents $m$ has been set to two (for the reasons explained in \cref{subsubsec:qubo-size-n-penalty-vals}) and, as a consequence, the complexity of this step turns out to be $\mathcal{O}(n^3)$. It is also worth noticing that the empty set is a valid parent set.

After that, the $\alpha_{ijk}$ hyperparameters of the Dirichlet priors are set to the uninformative value $1/(r_i\cdot q_i)$, with $r_i$ being the number of states of the $i$-th variable and $q_i$ (denoted as $q_{i{\pi}}$ in the pseudocode) being the number of states of the considered parent set. In practice, all $\alpha_{ijk}$ related to a specific variable $i$ and parent set $\pi$ (denoted as $\alpha_{i{\pi}jk}$ in the pseudocode) have the same value; further details about this choice are provided in \cref{subsec:ogorman-implementation-res}. In this step, the $\alpha_{ijk}$ value for all possible "variable" - "parent set" - "parent set state" - "variable state" combinations must be generated; hence, the complexity is $\mathcal{O}(n^3r_{max}^3)$, where $r_{max}$ is the maximum number of states of the Bayesian variables.

\begin{algorithm}[b!]
  \footnotesize
  \SetKw{and}{and}

  \KwIn{number of Bayesian variables $n$, list of number of states $r$, list of parent sets $parentSets$, prior distributions hyperparameters $\alpha=(\alpha_{i\pi jk})$, dataset $examples$
  }
  \KwResult{$s = (\{s_i(\pi)\ s.t.\ \pi \in parentSets[i]\})_{i=1}^n$ with $s_i(\pi)$ being the score for the Bayesian variable $i$ given the parent set $\pi$}
  \SetKwProg{ffun}{function}{:}{}

  \ffun(\tcp*[f]{Eq.~\eqref{eq:s-calc}}){calcSi(i, $\pi$, r, $\alpha$, examples)}{
    $q_{i{\pi}} \gets \prod_{p \in \pi}r_p$\tcp*{$q_{i{\pi}}=1$ if $\pi = \emptyset$}
    $sum \gets 0$\;
    \For{$j \gets 1 $ \KwTo $q_{i{\pi}}$}{
      $\alpha_{i\pi j} \gets \sum_{k=1}^{r_i}{\alpha_{i\pi jk}}$\;
      $N_{i{\pi}j} \gets \sum_{k=1}^{r_i}{calcNi{\pi}jk(examples,\pi,i,j,k,r)}$\;
      $sum \gets sum + \ln\Gamma(\alpha_{i\pi j}) - \ln\Gamma(\alpha_{i\pi j} + N_{i{\pi}j})$\;
      \For{$k \gets 1$ \KwTo $r_i$}{
        $N_{i{\pi}jk} \gets calcNi{\pi}jk(examples,\pi,i,j,k,r)$\;
        $sum \gets sum + \ln\Gamma(\alpha_{i\pi jk} + N_{i{\pi}jk}) - \ln\Gamma(\alpha_{i\pi jk})$\;
      }
    }
    \Return{-sum}\;
  }

  \For{$i \gets 1$ \KwTo $n$}{
    \For{$\pi \in parentSets[i]$}{
      $s_{i}(\pi) \gets calcSi(i,\pi,r,\alpha,examples))$\;
    }
  }

  \Return{s}\;

  \caption{\it calcS(n, r, parentSets, $\alpha$, examples)}
  \label{alg:calcS}
\end{algorithm}

\begin{algorithm}[t!]
  \footnotesize

  \KwIn{number of Bayesian variables $n$, list of parent sets $parentSets$, score values $s$}
  \KwResult{$w = (\{w_i(\pi)\ s.t.\ \pi \in parentSets[i]\})_{i=1}^n$ with $w_i(\pi)$ being the weight calculated for the Bayesian variable $i$ given the parent set $\pi$}
  \SetKwProg{ffun}{function}{:}{}

  \ffun(\tcp*[f]{Eq.~\eqref{eq:w-formula}}){calcWi(i, $\pi$, s)}{
    \uIf{$\pi = \emptyset$}{
      \Return{$s_i(\emptyset)$\;}
    }\uElseIf{$size(\pi) = 1$}{
      \Return{$s_{i}(\pi) - s_i(\emptyset)$}\;
    }\ElseIf{$size(\pi) = 2$}{
      $p_{1}, \ p_{2} \gets \pi[1],\ \pi[2]$\;
      \Return{$s_{i}(\pi) - s_{i}(\{p_{1}\}) - s_{i}(\{p_{2}\}) + s_i(\emptyset)$}\;
    }
  }

  \For{$i \gets 1$ \KwTo $n$}{
    \For{$\pi \in parentSets[i]$}{
      $w_{i}(\pi) \gets calcWi(i,\pi,s)$\;
    }
  }
  \Return{w}\;

  \caption{\it calcW(n, parentSets, s)}
  \label{alg:calcW}
\end{algorithm}

The next step consists in computing the local scores $s$ for all possible "Bayesian variable" - "parent set" combinations according to Eq.~\eqref{eq:s-formula}. Nevertheless, due to the factorial nature of the $\Gamma$ function and the presence of multiplications of $\Gamma$ values, the calculations turn out to be feasible only for very small datasets; indeed, the values quickly go out of range. The solution lies in moving the logarithm inside through algebraic steps until its argument becomes the gamma function alone. In the implementation presented here, the natural logarithm ($\ln$) has been used, and the resulting formula, the one employed in \cref{alg:calcS}, is the following:
\begin{equation}
  \label{eq:s-calc}
  \begin{split}
      s_i(\varPi_i(B_s)) = - \displaystyle\sum_{j=1}^{q_i}\Bigl[\ln(\Gamma(\alpha_{ij})) - \ln(\Gamma(N_{ij} + \alpha_{ij})) + \displaystyle\sum_{k=1}^{r_i}[\ln(\Gamma(N_{ijk} + \alpha_{ijk})) - \ln(\Gamma(\alpha_{ijk}))]\Bigr].
  \end{split}
\end{equation}
This form allows exploiting the (natural) log-gamma function (denoted as $\ln\Gamma$ in the pseudocode) instead of the gamma one, a function characterised by a far slower growth. Moreover, by doing this, the products in Eq.~\eqref{eq:s-formula} are replaced by additions, further decreasing the risk of out of range values. Concerning the pseudocode, the $calcNi{\pi}jk$ procedure just computes the number of times the variable $i$ is in its $k$-th state while its parent set $\pi$ is in its $j$-th state (in the case of an empty parent set, the state of the $i$-th variable alone is considered). The complexity of the algorithm is $\mathcal{O}(n^3Nr_{max}^3)$.

Once the score values $s$ have been calculated, it is possible to compute the parent set weights $w$ for the score Hamiltonian according to \eqref{eq:w-formula}. The pseudocode for this step is available in \cref{alg:calcW}. As previously mentioned, the maximum number of allowed parents has been set to two, hence the pseudocode does not take into account cases with larger parent sets. The complexity of the algorithm for the computation of $w$ is $\mathcal{O}(n^3)$.

Eventually, the penalty values must be calculated, starting from the auxiliary quantities $\Delta$, which are computed according to Eq.~\eqref{eq:delta-formula} and \eqref{eq:delta-prime-formula} with a complexity of $\mathcal{O}(n^4)$; actually, this complexity derives from the data structure used in the code to store the parent sets (according to the formulas, the complexity would be $\mathcal{O}(n^3)$). Given $\Delta$, all penalties can be determined. In detail, $\delta_{max}^{(i)}$ is computed for each Bayesian variable according to \eqref{eq:delta-max-formula} with a resulting complexity (for all $\delta_{max}^{(i)}$) of $\mathcal{O}(n^2)$. Instead, the penalty bound related to the consistency Hamiltonian (Eq.~\eqref{eq:delta-consist-formula}) can be simplified due to independence of $\delta_{trans}$ from its superscript indices. The outcome is the following:
\begin{equation}
  \label{eq:delta-consist-simplified-calc}
  \delta_{consist} > (n-2)\delta_{trans}.
\end{equation}
In practice, $\delta_{trans}$ is computed according to Eq.~\eqref{eq:delta-trans-formula} with complexity $\mathcal{O}(n^2)$ (notice that $\delta_{trans}$ is a single value). Then, $\delta_{consist}$ is calculated according to the simplified bound (Eq.~\eqref{eq:delta-consist-simplified-calc}) with complexity $\mathcal{O}(1)$. In order to satisfy the lower bounds, the penalty values have been set to the boundary values plus one.

\begin{algorithm}[t!]
  \footnotesize
  \SetNoFillComment

  \KwIn{zero matrix $Q$, number of Bayesian variables $n$, list of parent sets $parentSets$, parent set weights $w$, list of penalty values $\delta_{max}$, penalty value $\delta_{trans}$, penalty value $\delta_{consist}$}
  \KwResult{QUBO matrix $Q$ filled according to $H_{score}$, $H_{max}$, $H_{trans}$, and $H_{consist}$}

  \For{$i \gets 1$ \KwTo $n$}{
    \BlankLine
    \tcc{$H_{score}$-related terms (Eq.\,(\ref{eq:score-hamiltonian-formula}))}
    \For{$\pi \in parentSets[i]$}{
        \uIf(\tcp*[f]{diagonal elements}){$size(\pi) = 1$}{
          $j \gets \pi[1]$\;
          $row \gets col \gets indexOf(d_{ji})$\;
          $Q[row][col] \gets Q[row][col] + w_{i}(\pi)$\;
        }\ElseIf(\tcp*[f]{out-of-diagonal elements}){$size(\pi) = 2$}{
          $x,\ y \gets \pi[1],\ \pi[2]$\;
          $row,\ col \gets indexOf(d_{xi}),\ indexOf(d_{yi})$\;
          $Q[row][col] \gets Q[row][col] + w_{i}(\pi)$\;
        }
    }

    \BlankLine
    \tcc{$H_{max}$-related terms (Eq.\,(\ref{eq:max-hamiltonian-formula}))}
    $m \gets 2$\tcp*{max.\,num.\,of parents}
    $sqBinVars \gets binaryVarsInSquare()$\tcp*{$d_i$ and $y_i$ in (\ref{eq:max-hamiltonian-formula}) are sums of binary vars}
    $c \gets binaryVarsCoefficientsInSquare()$\tcp*{the coefficients are either -1 or -2}
    \For{$j \gets 1$ \KwTo $size(sqBinVars)$}{
      $row \gets col \gets indexOf(sqBinVars[j])$\tcp*{diagonal elements indices}
      $Q[row][col] \gets Q[row][col] + \delta_{max}^{(i)} \cdot c[j]^2$\tcp*{squared term}
      $Q[row][col] \gets Q[row][col] + \delta_{max}^{(i)} \cdot (2 \cdot m \cdot c[j])$\tcp*{double product with $m$}
      \BlankLine
      \For(\tcp*[f]{out-of-diagonal elements}){$k \gets j+1$ \KwTo $size(sqBinVars)$}{
        $row \gets indexOf(sqBinVars[j])$\;
        $col \gets indexOf(sqBinVars[k])$\;
        $Q[row][col] \gets Q[row][col] + \delta_{max}^{(i)} \cdot (2 \cdot c[j] \cdot c[k])$\tcp*{double product between vars}
      }
    }

    \BlankLine
    \tcc{$H_{cycle}$-related terms}
    \For{$j \gets i+1$ \KwTo $n$}{
      \tcc{$H_{trans}$-related terms (Eq.\,(\ref{eq:trans-hamiltonian-formula}))}
      \For{$k \gets j+1$ \KwTo $n$}{
          $row \gets col \gets indexOf(r_{ik})$\;
          $Q[row][col] \gets Q[row][col] + \delta_{trans}$\tcp*{$r_{ik}$ coefficient (diagonal element)}
          $row,\ col \gets indexOf(r_{ij}),\ indexOf(r_{jk})$\;
          $Q[row][col] \gets Q[row][col] + \delta_{trans}$\tcp*{$r_{ij} \cdot r_{jk}$ coefficient}
          $row,\ col \gets indexOf(r_{ij}),\ indexOf(r_{ik})$\;
          $Q[row][col] \gets Q[row][col] - \delta_{trans}$\tcp*{$r_{ij} \cdot r_{ik}$ coefficient}
          $row,\ col \gets indexOf(r_{ik}),\ indexOf(r_{jk})$\;
          $Q[row][col] \gets Q[row][col] - \delta_{trans}$\tcp*{$r_{ik} \cdot r_{jk}$ coefficient}
      }

      \BlankLine
      \tcc{$H_{consist}$-related terms (Eq.\,(\ref{eq:consist-hamiltonian-formula})}
      $row,\ col \gets indexOf(d_{ji}),\ indexOf(r_{ij})$\;
      $Q[row][col] \gets Q[row][col] + \delta_{consist}$\tcp*{$d_{ji} \cdot r_{ij}$ coefficient}
      $row \gets col \gets indexOf(d_{ij})$\;
      $Q[row][col] \gets Q[row][col] + \delta_{consist}$\tcp*{$d_{ij}$ coefficient (diagonal element)}
      $row,\ col \gets indexOf(d_{ij}),\ indexOf(r_{ij})$\;
      $Q[row][col] \gets Q[row][col] - \delta_{consist}$\tcp*{$d_{ij} \cdot r_{ij}$ coefficient}
    }
  }

  \Return{Q}\;

  \caption{\it fillQ(Q, n, parentSets, w, $\delta_{max}$, $\delta_{trans}$, $\delta_{consist}$)}
  \label{alg:fillQ}
\end{algorithm}

At this point, it is possible to fill the QUBO matrix $Q$ as shown in \cref{alg:fillQ}; the matrix, whose size has been already described in \cref{subsubsec:qubo-size-n-penalty-vals}, initially contains only zeros (see line 9 of \cref{alg:calcQUBOMatrix}, whose complexity is $O(n^4)$). In detail, the first section of \cref{alg:fillQ} (lines 2-12) is related to the score Hamiltonian, namely, to Eq.~\eqref{eq:score-hamiltonian-formula}. For each "Bayesian variable" - "parent set" combination, the parent set weight $w_i(\pi)$ ($w_i(J)$ in O'Gorman's formulation) is added to the appropriate cell; the outermost loop, which includes almost all the pseudocode, is the one iterating on the Bayesian variables. In practice, the coefficients of the linear terms of Eq.~\eqref{eq:score-hamiltonian-formula}, i.e., the terms involving only one QUBO variable ($d_{ji}$), are summed to cells of $Q$ located on the main diagonal. Instead, the coefficients of the quadratic terms, which involve two QUBO variables ($d_{xi}d_{yi}$), contribute to cells outside the diagonal; indeed, the first variable determines the row, while the other the column. The subsequent part of the algorithm is related to the max Hamiltonian (lines 13-25), i.e., to Eq.~\eqref{eq:max-hamiltonian-formula}. The approach used for the coefficient insertion is similar to that employed for the score Hamiltonian, however the presence of a square must be taken into account. Hence, for each Bayesian variable (outermost loop), the binary variables involved in Eq.~\eqref{eq:max-hamiltonian-formula} and their coefficients inside the square are determined and stored in two lists (lines 14-15). Then, based on the square expansion, the resulting linear and quadratic coefficients (which include the multiplication by $\delta_{max}^{(i)}$) are summed to the respective cells. Finally, there is the section related to the transitivity and consistency Hamiltonians (lines 26-43), namely, to Eq.~\eqref{eq:trans-hamiltonian-formula} and \eqref{eq:consist-hamiltonian-formula}. In this case, due to the small amount of terms in the formulas and the absence of squares, the procedure is simpler. In detail, lines 28-35 sum the coefficients given by Eq.~\eqref{eq:trans-hamiltonian-formula} to the corresponding locations for each set of three Bayesian variables ($H_{trans}$ penalties). Analogously, lines 37-42 add the coefficients given by Eq.~\eqref{eq:consist-hamiltonian-formula} to the appropriate cells for any pair of Bayesian variables ($H_{consist}$ penalties). The resulting complexity for the matrix filling procedure (\cref{alg:fillQ}) is $\mathcal{O}(n^3)$.

It is also worth mentioning that, in the QUBO matrix $Q$, the variables are ordered in the following way: first, the binary variables encoding the edges ($d_{ij}$); then, the binary slack variables ($y_{il}$) related to the maximum parent constraint; finally, the binary variables ($r_{ij}$) encoding the topological order. By sorting appropriately the variables in the quadratic terms (they define the row and column indices), the outcome is an upper triangular matrix; otherwise, the content of the non-zero cells below the main diagonal should be transferred to the corresponding cells above the diagonal (summing up the values).

\subsection{Complexity}
\label{subsec:complexity}
The overall complexity of the QUBO matrix construction (Algorithm \ref{alg:calcQUBOMatrix}) is $\mathcal{O}(n^4 + n^3Nr_{max}^3)$. Hence, it is determined by several factors: the number of Bayesian variables ($n$) of the considered BNSL problem, the number of examples ($N$) in the dataset, and the maximum number of states ($r_{max}$) among the Bayesian variables. In particular, if
the number of Bayesian variables is smaller than the dataset size, the dominant complexity term becomes $n^3Nr_{max}^3$. The situation just depicted is typical. Indeed, $n$ cannot be too big due to the the limitations (in the number of qubits and connectivity) of the current quantum annealers, whereas $N$ must be considerably large to provide enough information to learn from. Therefore, typically, the calculation of the local score values $s$ (Algorithm \ref{alg:calcS}) turns out to be the most expensive operation in the QUBO matrix construction; otherwise, it would be the initialization of $Q$ to zero (on a par with the $\Delta$ calculation, given the implementation of the operation in question). Concerning the maximum number of states of the Bayesian variables, its contribution is particularly relevant for Bayesian variables with continuous states. In fact, the variables in question must be discretized, and the greater the representation accuracy, the higher the execution time.

\subsection{Execution speedup}
\label{subsec:exec-speedup}
The construction of the QUBO matrix (including the intermediate values calculation) is what takes most of the execution time. In general, a speedup could be obtained by using a different programming language such as C++; nevertheless, this is not feasible here due to D-Wave's Ocean library, which is necessary to interface with the quantum annealer and is written in Python. Instead, a valid solution consists in performing a dynamic compilation of the code through Numba~\cite{numba}, a just-in-time compiler for Python. In detail, Numba requires to apply a decorator to the functions that must be compiled. Then, during the execution, the first time a function with a decorator is called, it is compiled into machine code, and all the subsequent calls will run the machine code instead of the original Python code. It is important to notice that Numba works better on code including loops, NumPy arrays and library functions. Moreover, the speedup is effective only if a function is called several times; otherwise, in the case of one call in a run, the execution will be slower.

Actually, Numba has been exploited only in the experiments related to the divide et impera approach (\cref{subsec:divide-et-impera-res}), since it has been introduced after the completion of the experiments related to O'Gorman's algorithm (\cref{subsec:ogorman-implementation-res}). It is also worth mentioning that the dynamic compilation has been applied only to the two functions called most often in the execution (i.e., $calcNi{\pi}jk$ and another internal procedure).

\section{Divide et Impera Approach}
\label{sec:divide-et-impera}
\noindent
Embedding problems in the QPU topology requires a huge number of qubits due to the limited connectivity of the current quantum annealers. Moreover, the QUBO formulation of the BNSL problem proposed by O’Gorman et al. is densely connected by definition, making infeasible its application even to instances with a not-so-high number of variables. For these reasons, a divide et impera approach has been developed and tested; the pseudocode is shown in \cref{alg:divide-et-impera}.

\begin{algorithm}[ht!]
    \footnotesize
    \SetNoFillComment

    \KwIn{number of variables of the original BNSL problem $n$, number of variables for each subproblem $k$, list of number of states $r$, dataset $examples$}
    \KwOut{adjacency matrix of the solution to the original problem $sol$}

    \BlankLine
    \tcc{Subproblems formulation}
    $subproblems \gets combinations(n, k)$\;
    $subproblemsR \gets filter(r, subproblems)$\;
    $subproblemsEx \gets filter(examples, subproblems)$\;

    \BlankLine
    \tcc{Subproblems solution}
    $S \gets Set()$\;
    \For{$i \gets 0$ \KwTo $size(subproblems)$}{
        $subprob,\,subprobR,\,subprobEx \gets subproblems[i],\,subproblemsR[i],\,subproblemsEx[i]$\;
        $subprobQ \gets calcQUBOMatrix(size(subprob),subprobR,subprobEx)$\tcp*{\cref{alg:calcQUBOMatrix}}
        $subprobAdjMatrix \gets solveQUBO(subprobQ)$\;
        $S.add(subprob, subprobAdjMatrix)$\;
    }

    \BlankLine
    \tcc{Original solution reconstruction}
    $C,\ P \gets countEdgesAndPenalties(S, k)$\;
    $sol \gets zeroMatrix()$\;

    \For{$i \gets 0$ \KwTo $n$}{
        \For{$j \gets 0$ \KwTo $n$}{
            \If{$i \neq j$}{
                \If(\tcp*[f]{first strategy's condition: $C_{ij} > C_{ji}$}){$(C_{ij} - P_{ij}) > 0$}{
                    $sol[i][j] = 1$\;
                }
            }
        }
    }

    \Return{$sol$}\;
    \caption{\it{divideEtImpera(S, k, N)}}
    \label{alg:divide-et-impera}
\end{algorithm}

The first step is the subproblems formulation (lines 1-3). Let $n$ be the number of variables of the original BNSL problem, $r$ be an array containing the number of states for each variable, and $examples$ be a $N \times n$ matrix representing the dataset. The BNSL subproblems are generated as combinations of the $n$ variables taken $k$ at a time, where $k$ represents the desired number of variables for each subproblem. In detail, all possible combinations of variables are generated, and each subproblem is identified by the ($examples$ matrix column) indices of the variables included. In practice, the \textit{combinations} function from the Python \textit{itertools} module is used. The complexity of this procedure is $O(c\binom{n}{k})$, where $c$ is the (constant) cost for creating a list of indices. It is also worth mentioning that $k$ should be larger or equal than 3 since that is the minimum reasonable number of variables for the QUBO encoding ($H_{trans}$ assumes $n \ge 3$). Then, for each $k$-variables combination, it is necessary to filter the $r$ vector and the $examples$ matrix, obtaining a vector of $k$ elements and a $N \times k$ matrix, respectively. The total complexity of the filtering operations is $O(\binom{n}{k}(k + Nk))$.

After the subproblems generation, the implementation of O'Gorman's algorithm presented in \cref{sec:ogorman-implementation} can be applied to each subproblem (line 7), obtaining the respective QUBO matrices, which can be submitted to the annealer or solved with alternative methods (line 8). The outcome is an adjacency matrix for each subproblem. Regarding the complexity of this step, it is linear with respect to the number of subproblems ($\binom{n}{k}$).

Eventually, the solution to the original BNSL problem must be reconstructed starting from the subproblems solutions (lines 11-21). Let $S$ be the set of all subproblems solutions, where each solution consists of the list of indices of the variables included and an adjacency matrix for the corresponding graph. The relevant information here is the presence of edges, thus the first step consists in counting how many times each edge appears in the subproblems solutions. Indeed, each pair of variables is present in more than one subproblem. Let $C$ be the set of counts, with $C_{ij}$ representing the number of appearances of the $(i,j)$ edge (the edges are directed, hence $(i,j)$ and $(j,i)$ are different). After the counting phase, whose complexity is $O(\binom{n}{k}k^2)$, the reconstruction of the solution can start. Actually, two strategies have been developed for this. The first one (the simplest one) consists in inserting in the adjacency matrix of the original problem every edge $(i,j)$ that appears at least one time in one subproblem, i.e., for which $C_{ij} > 0$. If both $C_{ij}$ and $C_{ji}$ are larger than 0, then the edge with the highest number of counts is picked; in this way, cycles of two nodes are avoided (the resulting graph must be a DAG). Instead, the second strategy (the one shown in the pseudocode) requires additional information to perform the reconstruction, namely, the penalty values $P_ {ij}$. Basically, $P_{ij}$ represents the number of subproblems including the variables $i$ and $j$ in which the edge $(i,j)$ is not present; hence, the computation complexity is the same as for $C$. In practice, an edge $(i,j)$ is added to the final solution if the difference between $C_{ij}$ and $P_{ij}$ is larger than 0, otherwise is discarded. Note that if this condition is satisfied, so are those of the first method ($C_{ij} > 0$ and $C_{ij} > C_{ji}$). In the experiments presented in the next section, only the second strategy has been exploited, since some preliminary experiments have confirmed its superiority. Concerning the resulting complexity of the reconstruction phase, it is $O(\binom{n}{k}k^2 + n^2)$.

\section{Empirical Evaluation}
\label{sec:empirical-evaluation}
\noindent
This section deals with the Bayesian problems selected, the datasets generation procedure employed, the methods tested, the experimental setup, and the results obtained for both O'Gorman's algorithm and the divide et impera approach.

\subsection{Bayesian problems}
\label{subsec:problems}
\noindent
Three out of the four Bayesian problems used have been selected from the examples provided by the Bayes Server site \cite{bayesserver}, whereas the last one (the Lung Cancer) has been taken from a different source \cite{lung_cancer_source}. In detail, the implementation of O'Gorman's algorithm has been tested on the Monty Hall, the Lung Cancer and the Waste problem. Instead, the divide and impera approach has been tested on the Lung Cancer, the Waste, and the Alarm problem. It is also worth highlighting that most of these problems have been subjected to some modifications (explained in the following paragraphs) before applying the presented methods.

\begin{figure}[t!]
  \centering
  \begin{tabular}{ll}
    \includegraphics[width=0.5\linewidth]{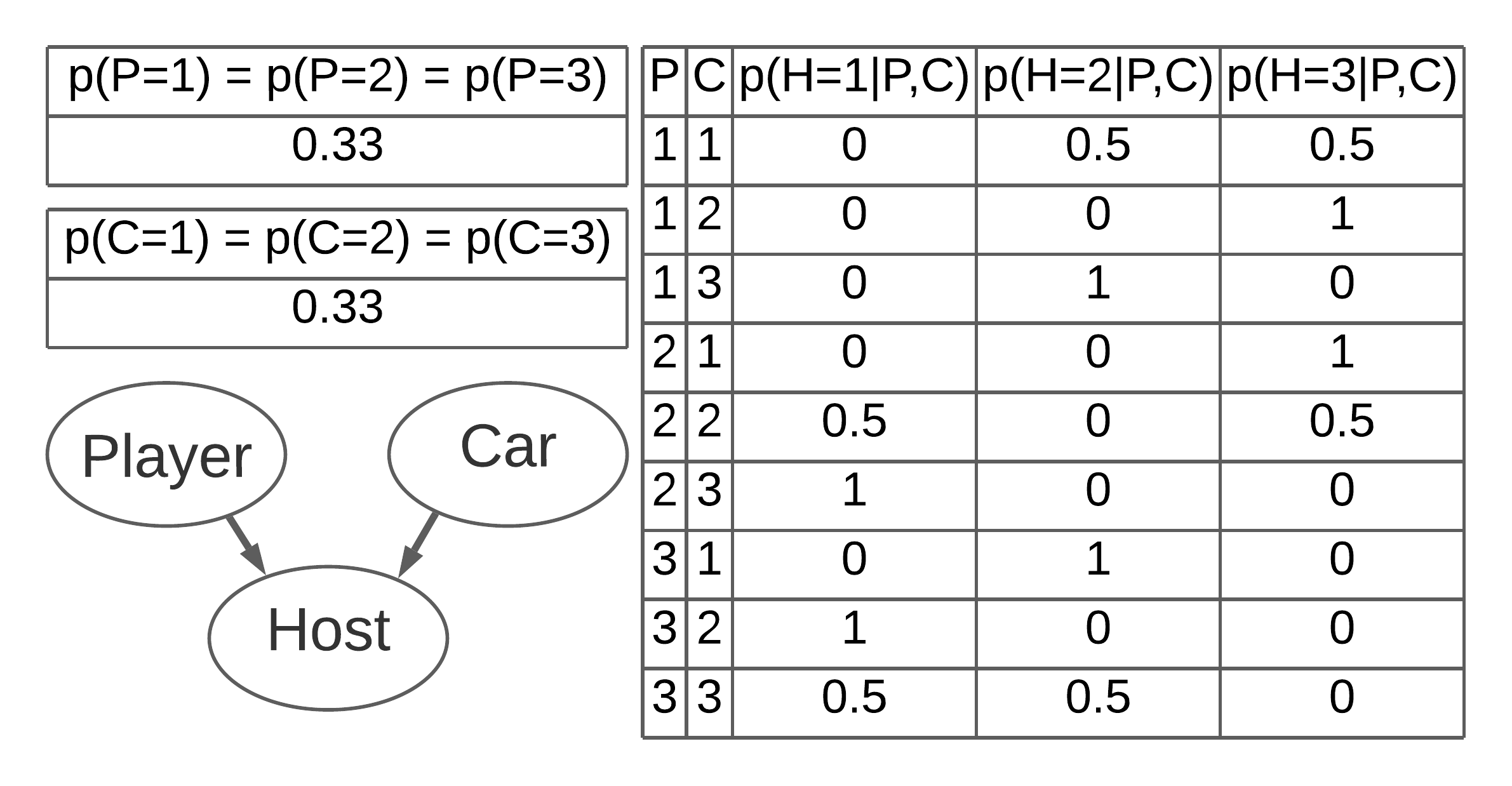} & \includegraphics[width=0.4\linewidth]{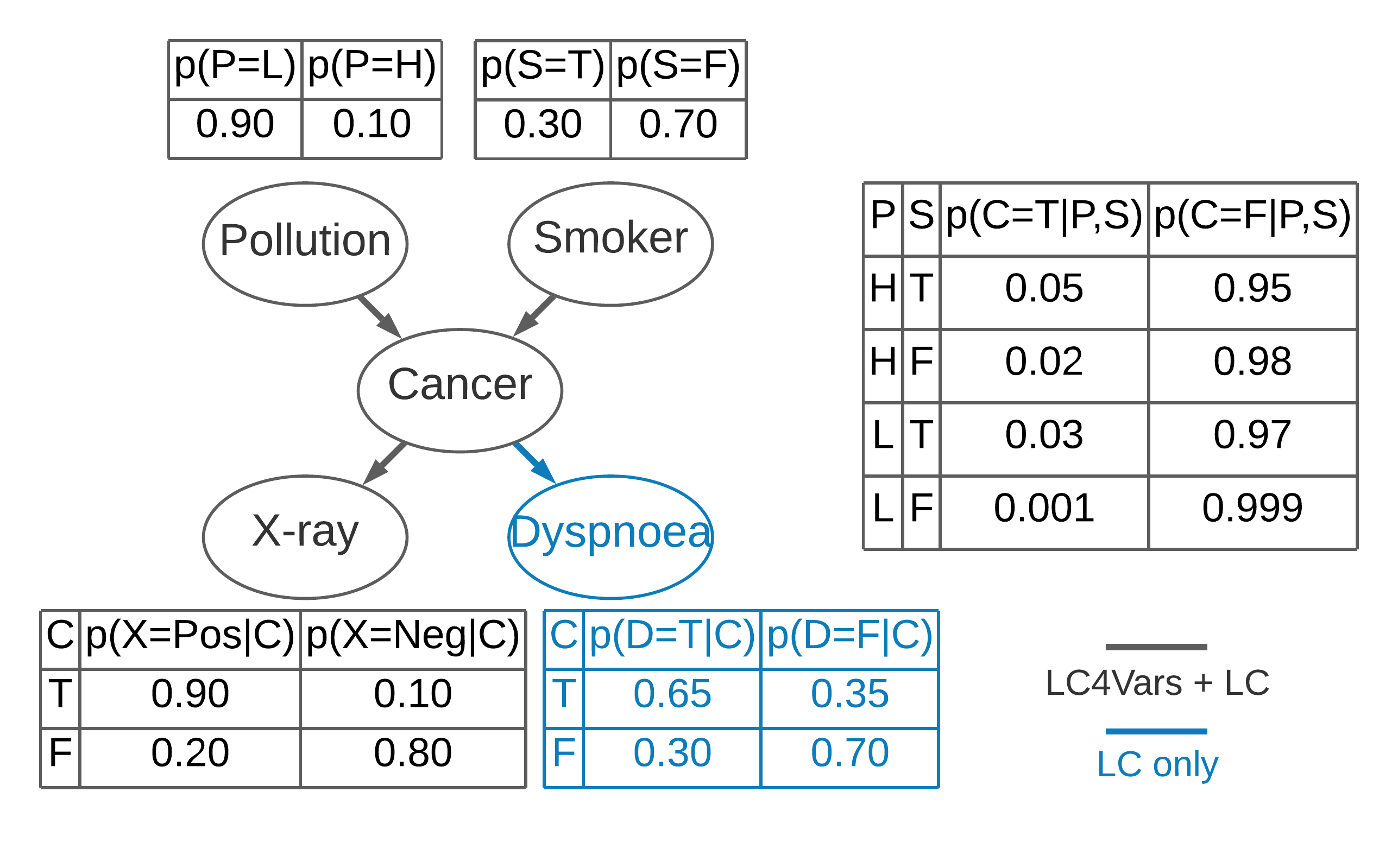}\\
  \end{tabular}
  \fcaption{Monty Hall Problem (left) and Lung Cancer (right).}
  \label{fig:MHP_LC}
\end{figure}

\paragraph{Monty Hall Problem}
The Monty Hall Problem (MHP) has been chosen because of its simplicity. In detail, the Bayesian network of the problem (see \cref{fig:MHP_LC}, on the left) is composed of three variables ($n=3$), and each variable has three possible states ($\{1,2,3\}$). Actually, three is also the minimum reasonable number of variables for O'Gorman's QUBO formulation. Indeed, the transitivity Hamiltonian is based on the assumption that at least three Bayesian variables are present.

\begin{figure}[b!]
  \centering
  \includegraphics[width=\linewidth]{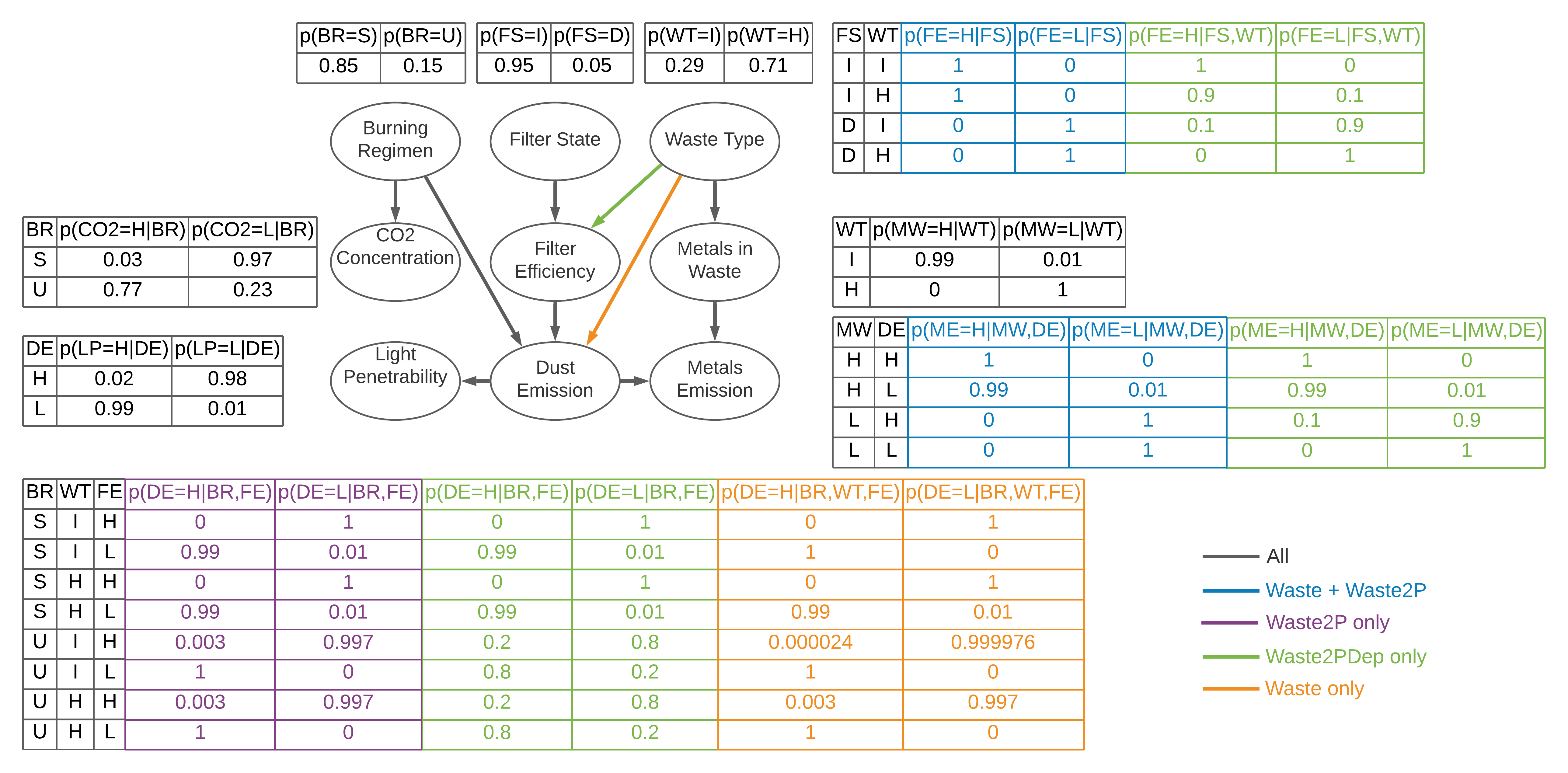}
  \fcaption{Waste.}
  \label{fig:Waste}
\end{figure}

\paragraph{Lung Cancer}
The Lung Cancer problem has been selected due to its (not excessively) higher number of variables with respect to the MHP. In particular, the original network (LC) consists of $n=5$ variables and each Bayesian variable admits two possible states. Actually, also a variant (LC4Vars) with $n=4$ variables obtained by removing the "Dyspnoea" node has been tested here. In this way, a more accurate analysis on the problem size scaling could be performed. Both networks are shown in \cref{fig:MHP_LC} (right).

\paragraph{Waste}
The Waste problem has been chosen for different reasons: it has a considerably larger size than the previous two; it includes continuous Bayesian variables, and also a variable with more than two parents. In detail, the Bayesian network of the original problem is composed of nine variables ($n=9$), of which three are discrete with two states, and six are continuous. Since the QUBO encoding admits only discrete variables, a discretization has been required for the continuous ones. However, it is not possible to use a single discretization threshold because the variables have different mean values. Hence, each continuous variable has been transformed into a discrete one with two states by applying the following procedure: first, the lowest and the highest values are identified by evaluating all the settings of the parent variables (this is possible since the number of variables and states are small), and the average of the two values is kept as a discretization threshold; then, for each combination of parent states, the mean and the variance of the continuous variable are taken down, and the probability of the Gaussian with these parameters having a value higher than the threshold is set as the probability of the high state ($H$, opposed to the low state $L$). Specifically, to determine the lowest (highest) value, the respective variance is subtracted from (summed to) the minimum (maximum) mean value observed; moreover, in the case of a continuous variable with a continuous parent, the parent is discretized first, and the lowest and highest values are used as evidence for the parent states $L$ and $H$. The resulting Bayesian problem is denoted here as Waste.

\cref{fig:Waste} summarizes all the considered variants of the problem. In particular, Waste differs from the original problem only in the lack of the edge between \textit{Waste Type} and \textit{Filter Efficiency}, which has been lost in the discretization procedure. Instead, in Waste2P (variant of Waste), the edge between \textit{Waste Type} and \textit{Dust Emission} has been removed so that the maximum number of parents is equal to two (in practice, the most balanced probability values have been kept for the \textit{Dust Emission} node). Finally, Waste2PDep is a variant of Waste2P in which the edge between \textit{Waste Type} and \textit{Filter Efficiency} has been reintroduced by manually altering some probability values (in the \textit{Filter Efficiency} probability table). In addition, in Waste2PDep, some other probabilities have been slightly changed (\textit{Dust Emission} and \textit{Metals Emission} tables) in order to have more balanced probability distributions.

\begin{sidewaysfigure}
  \centering
  \includegraphics[width=\linewidth]{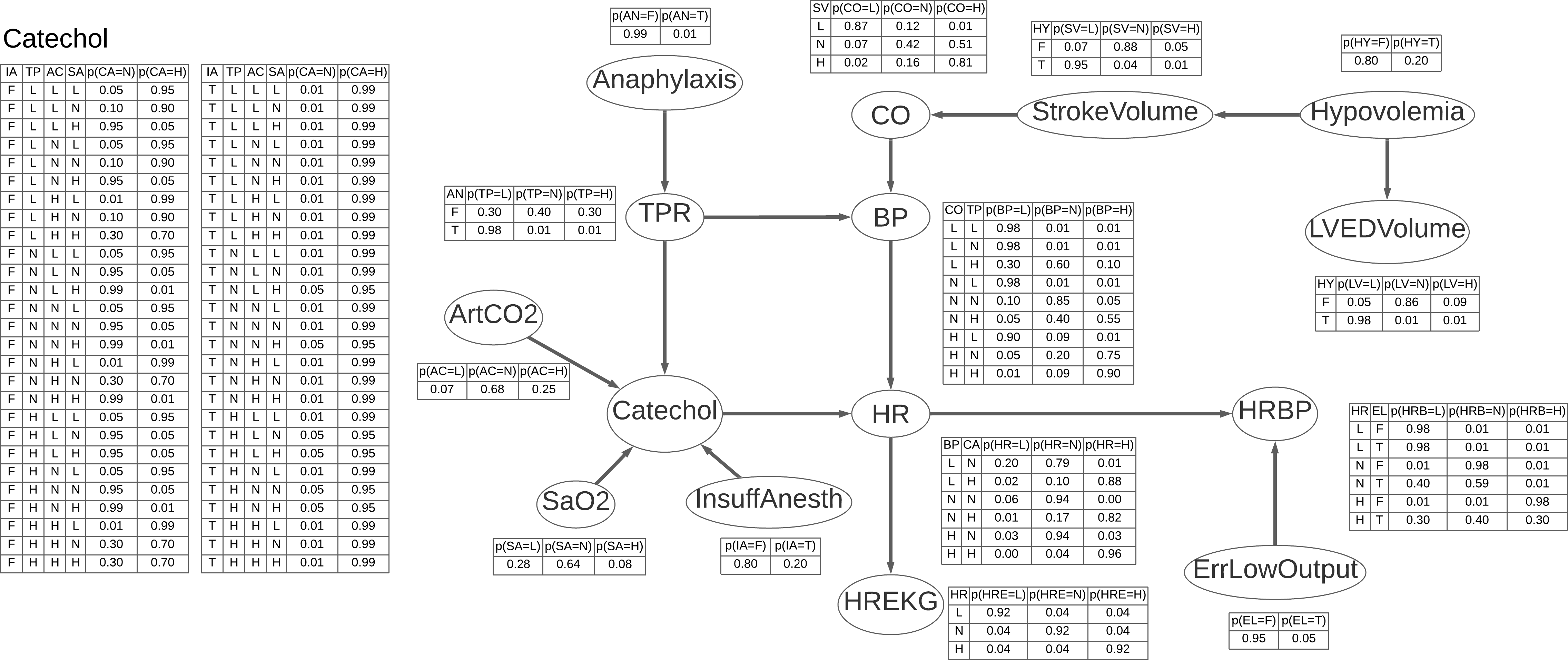}
  \fcaption{Alarm.}
  \label{fig:Alarm}
\end{sidewaysfigure}

\paragraph{Alarm}
Eventually, Alarm has been picked mainly for its size. Indeed, it is quite close to the maximum BNSL problem size that can be embedded in the Pegasus topology using O'Gorman's formulation ($\approx 18$). Moreover, the presence of a variable with four parents allows evaluating the ability of the divide et impera approach to reconstruct Bayesian networks with more than two parents for a single variable. Actually, the original Alarm problem consists of 38 Bayesian variables, whereas the version used here includes only 15 of them (with their structure preserved). The Bayesian network employed in the experiments is shown in \cref{fig:Alarm}.

\subsection{Datasets generation}
\label{subsec:datasets-gen}
For each problem, several datasets have been generated varying both the size and the creation method. Specifically, three dataset sizes ($N$) have been used, namely, $10^4$ (10K), $10^5$ (100K), and $10^6$ (1M). Regarding the generation methods, two have been employed. The first one consists in generating $N$ examples through the \textit{uniform} function from the Python \textit{random} module by sampling from the network probability distribution. Instead, the second method aims at generating datasets with zero variance, i.e., with combinations of states appearing exactly the expected amount of times. In particular, in order to generate the expected datasets, the probability $p$ of every combination of states of the network variables is calculated; then, for each combination, $\lfloor N*p \rfloor$ examples are inserted in the dataset. Actually, for some probability values, it may not be possible to generate an integer number of examples, and, consequently, the variance of the dataset is not exactly zero. In addition, the resulting dataset may have a number of samples lower than the desired one. For instance, for the Alarm problem, the dataset of size $N = 10^4$ generated using this method has a considerably lower number of samples ($\approx 9000$) due to the presence of many state combinations that are not represented at all because of their very low probability values (the other problem datasets are not significantly affected by this issue). The datasets generated using the second method are denoted as \textit{Exp}.

\subsection{Methods and experimental setup}
\label{subsec:methods-n-setup}
After the application of O'Gorman's algorithm, the BNSL problem encoded in the QUBO matrix must be solved. For this purpose, three methods have been exploited in the experiments: quantum annealing (QA), simulated annealing (SA) and exhaustive search (ES). The functioning of quantum annealing has been already explained in \cref{subsec:quantum-annealing-and-dwave}. Instead, simulated annealing is a well-known classical metaheuristic technique used for solving optimization problems \cite{simulated_annealing}; further details about the algorithm can be found here~\fnm{b}\fnt{b}{ \url{https://en.wikipedia.org/wiki/Simulated_annealing}}. Eventually, the exhaustive search represents an optimized brute force approach.

In detail, a simple brute force on the QUBO representation would be unfeasible even for small BNSL instances due to the high number of binary variables. Hence, ES limits the brute force to the edge binary variables $d_{ij}$, while considering only the best setup of $y_{il}$ and $r_{ij}$. In particular, for each Bayesian variable $i$, the $y_{il}$ binary variables must be set so that $y_i$ (in Eq.~\eqref{eq:max-hamiltonian-formula}) is equal to the difference between $m$ and $d_i$. In this way, there is no penalty from the max Hamiltonian. Obviously, this is possible only for nodes that do not have more than $m$ parents; otherwise, the minimum penalty is given by $y_i$ equal to zero. Instead, setting the $r_{ij}$ binary variables is more complex. Indeed, computing the topological order of the graph encoded in $d_{ij}$ is not enough since, if two Bayesian variables $i$ and $j$ are not connected (i.e., it does not exist a path from one variable to the other), there is no straightforward setup for $r_{ij}$. In addition, setting all the uncertain binary variables to either 0 or 1 does not solve the problem since a cycle might be produced, with consequent penalty from $H_{trans}$. The solution consists in completing the graph encoded in $d_{ij}$ by adding one edge at a time (while verifying that no cycle is formed), and then computing the topological order of the resulting graph. In this way, it is possible to properly set all $r_{ij}$ binary variables and avoid any penalty from the cycle Hamiltonian. Obviously, if the graph encoded in $d_{ij}$ contains a cycle, it is not possible to avoid the penalty. The resulting complexity for setting $y_{il}$ and $r_{ij}$ turns out to be $\mathcal{O}(n^3)$ for sparse graphs and $\mathcal{O}(n^4)$ for dense graphs. In the end, these optimizations do not change the complexity class of ES with respect to the simple brute force (it remains exponential in the number of Bayesian variables $n$) but significantly reduce the number of operations to be performed. Another improvement that has been introduced in ES is the parallelization of the solutions evaluation.

Regarding the setup for the experiments on the implementation of O'Gorman's algorithm, different combinations of annealing parameters (number of annealer reads and annealing time) \cite{solver_params} have been tested for QA; the specific values are reported in the various results sections. In addition, the default annealing schedule has been employed (the system used is Advantage 1.1), and the QPU embedding has been performed through the \textit{EmbeddingComposite} class (see \cref{subsec:qpu-embedding}). As concerns SA, the implementation provided by D-Wave \cite{dwave_sim_annealing} has been exploited. Except for the number of reads (different values have been employed), the default configuration has been used, and the execution has been carried out on a local machine. Eventually, ES does not require to set parameters and has also been executed locally. In detail, a machine with a quad-core CPU (Intel i5-6400) and 16 GB of RAM has been used for the datasets generation, the QUBO matrices construction, and the execution of the classical methods.

Concerning the divide et impera approach, only one configuration of annealing parameters has been evaluated for QA; the reason and the values are specified in the related results section. Furthermore, the default annealing schedule has been employed, but, in this case, the system used is Advantage 4.1 since the previous model had already been dismissed. Instead, the QPU embedding method and the SA implementation are the same as those exploited in the experiments on O'Gorman's algorithm; the only difference lies in the number of reads used for SA, which has been set to a single value too. Finally, ES has not been evaluated on this approach due to its high time requirements. All the classical operations for the divide et impera approach have been executed on a machine with a quad-core CPU (Intel i7-7700HQ) and 16 GB of RAM.

Eventually, it is worth mentioning that, for both the implementation of O'Gorman's algorithm and the divide et impera approach, the performance-related analyses presented here involve only \textit{Exp} datasets because some preliminary tests have confirmed that the difference in terms of performance when working on \textit{Exp} or non-\textit{Exp} datasets is negligible.

\subsection{O'Gorman's algorithm results}
\label{subsec:ogorman-implementation-res}
Several experiments have been performed on the implementation of O'Gorman's algorithm. The results obtained are presented in the subsequent paragraphs.

\paragraph{QUBO formulation correctness and $\alpha_{ijk}$ hyperparameters}
The QUBO encoding must accurately represent the original BNSL problem; to this end, the $\alpha_{ijk}$ hyperparameters must be set appropriately. In detail, if $\alpha_{ijk}$ have suitable values, the image ($x^TQx$) of the expected solution will be the global minimum. To verify this, it is necessary to find the global minimum solution through ES and compare it with the expected one. In particular, the QUBO version of the expected solution ($x$) is obtained as follows: the Bayesian network that has been used to generate the dataset is exploited to set the $d_{ij}$ binary variables encoding the edges, whereas the best setup of $y_{il}$ and $r_{ij}$ is found using the same approach employed by ES (see \cref{subsec:methods-n-setup}).

Since the objective is to learn the structure of a Bayesian network from a set of data, the values of $\alpha_{ijk}$ must be uninformative, i.e., they must not encode information about the target Bayesian network. For this reason, the first values of $\alpha_{ijk}$ that have been tested are $N/(r_i\cdot q_i)$ and $1$, as proposed by Heckerman et al. \cite{heckerman}. The results obtained are reported in \cref{tab:alpha-values}. In detail, the first alternative ($N/(r_i\cdot q_i)$) has performed decently on the MHP problem (the smallest one), and extremely bad on all the others (characterised by a higher number of variables); in practice, $N$ prior counts uniformly distributed among all "variable" - "parent set" state combinations are added in Eq.~\eqref{eq:s-formula}. Instead, the second possibility (i.e., $1$) has not worked at all. Therefore, other values have been evaluated, namely, $1/(r_i \cdot q_i)$ and $1/r_i$, which have been selected with the idea of influencing the counts as little as possible. As reported in the same table, both of them have shown the desired behaviour, and the first one has been chosen as the default setup.

In particular, the ratios in \cref{tab:alpha-values} have been obtained using 8 datasets for each "problem" - "$\alpha_{ijk}$ value" combination; the datasets in question have been created with different sizes and exploiting both generation methods (see \cref{subsec:datasets-gen}). Specifically, four datasets of size $N=10^4$ (of which one \textit{Exp}), two datasets of size $N=10^5$ (of which one \textit{Exp}), and two datasets of size $N=10^6$ (of which one \textit{Exp}) have been used.

\begin{table}[ht]
  \tcaption{Ratio of the number of times in which the best and the expected solutions coincide to the number of tests (8 for each cell), for different problems and $\alpha_{ijk}$ values.}
  \label{tab:alpha-values}
  \centerline{\footnotesize\begin{tabular}{|c|c c c c|} \hline
    Problem & $\alpha_{ijk} = N/(r_i \cdot q_i)$ & $\alpha_{ijk} = 1$ & $\alpha_{ijk} = 1/(r_i \cdot q_i)$ & $\alpha_{ijk} = 1/r_i$ \\ \hline
    MHP     & 0.75 & 0.00 & \textbf{1.00} & 1.00 \\
    LC4Vars & 0.00 & 0.00 & \textbf{1.00} & 1.00 \\
    LC      & 0.00 & 0.00 & \textbf{1.00} & 1.00 \\ \hline
  \end{tabular}}
\end{table}

\paragraph{Dataset size and QUBO matrix construction time}
After the choice of the $\alpha_{ijk}$ value, the impact of the dataset size on the QUBO matrix construction time has been analyzed. As shown in \cref{tab:qubo-construction-time}, the required time increases linearly with the dataset size in accordance with the complexity $\mathcal{O}(n^4 + n^3Nr_{max}^3)$ discussed in \cref{subsec:complexity}. In practice, large dataset sizes turn out to be prohibitive, especially when constructing the QUBO matrix for problems with a high number of variables. Because of this and the fact that a dataset size larger than $N=10^4$ leads to no improvement in terms of performance (see \cref{tab:dataset-size-effect-on-performance}), it is advantageous to keep the dataset size limited.

Regarding the tables data, the time values in \cref{tab:qubo-construction-time} have been obtained using one \textit{Exp} and four non-\textit{Exp} datasets for each "problem" - "dataset size" combination (hence, five runs for each entry). Instead, the values in \cref{tab:dataset-size-effect-on-performance} have been acquired through QA, using \textit{Exp} datasets only, $10^4$ reads, 20$\mu$s of annealing time, and 10 runs for each dataset size. The metric reported in the second table is described in the performance section later on; for the time being, it is sufficient to know that larger values in the table correspond to better performance.

\begin{table}[ht]
  \tcaption{Average QUBO matrix ($Q$) construction time in seconds, for different problems and dataset sizes. For each entry, 5 different datasets (of which one \textit{Exp}) have been used.}
  \label{tab:qubo-construction-time}
  \centerline{\footnotesize\begin{tabular}{|c|c c c|}
    \hline
    Problem & $N = 10^4$ & $N = 10^5$ & $N = 10^6$ \\
    \hline
    MHP        &    0.55 &    4.03 &   40.03 \\
    LC4Vars    &    0.64 &    5.86 &   58.78 \\
    LC         &    1.40 &   13.55 &  136.21 \\
    Waste      &    9.79 &   98.85 & 1010.30 \\
    Waste2P    &    9.74 &   98.78 & 1001.96 \\
    Waste2PDep &    9.77 &  100.26 & 1007.61 \\
    \hline
  \end{tabular}}
\end{table}

\begin{table}[ht]
  \tcaption{Average solution value found by QA in Waste2PDep on \textit{Exp} datasets of different sizes. $10^4$ reads, 20$\mu$s of annealing time, and 10 runs (for each dataset size) have been used.}
  \label{tab:dataset-size-effect-on-performance}
  \centerline{\footnotesize\begin{tabular}{|c|c c c|}
    \hline
    Problem & $N = 10^4$ & $N = 10^5$ & $N = 10^6$ \\
    \hline
    Waste2PDep  & 0.963 & 0.960 & 0.965 \\
    \hline
  \end{tabular}}
\end{table}

\paragraph{Number of reads and annealing time (QA)}
The number of reads, i.e., measurements, and the annealing time per read are two extremely relevant parameters for the performance of QA. Hence, an extensive experimental evaluation has been performed, with 10 runs for each configuration; the results are reported in \cref{tab:num-reads-vs-annealing-time}. In detail, the maximum allowed number of reads (on D-Wave systems) is $10^4$, whereas the maximum annealing time per read is $2000 \mu$s. However, there is also an internal constraint that prevents an annealing time larger than $999\mu$s with $10^3$ reads, and an annealing time larger than $99\mu$s with $10^4$ reads. Looking at the results, it is clear that a higher number of reads provides better performance, and the same holds for the annealing time. Nevertheless, the number of reads has a more significant impact. Indeed, the results achieved with the maximum allowed number of reads and an annealing time far lower than the joint limit are clearly better than those achieved with the annealing time maximized. Hence, the best setup corresponds to the maximum allowed number of reads ($10^4$) and the joint limit annealing time ($99\mu$s).

\begin{table}[ht]
  \tcaption{Ratio of the number of times the global minimum is found by QA to the number of experiments (10 for each configuration), for different numbers of reads and annealing times on the LC \textit{Exp} dataset with size $N=10^4$.}
  \label{tab:num-reads-vs-annealing-time}
  \centerline{\footnotesize\begin{tabular}{|c|c c c c c c|}
    \hline
    & \multicolumn{6}{c|}{Annealing time ($\mu$s)} \\
    \hline
    reads \# & 1 & 10 & 20 & 50 & 99 & 999 \\
    \hline
    $10^3$    & ---  & 0.00 & ---  & ---  & ---  & 0.10 \\
    $10^4$    & 0.00 & 0.20 & 0.20 & 0.40 & 0.50 & ---  \\
    \hline
  \end{tabular}}
\end{table}

\paragraph{Performance}
Finally, all the methods described in \cref{subsec:methods-n-setup} have been applied to all the problems presented in \cref{subsec:problems} with the purpose of comparing their performance; the results are reported in \cref{tab:comp-performance}. In particular, for these experiments, \textit{Exp} datasets of size $N=10^4$ have been employed. Moreover, $10^4$ reads have been used for SA, whereas, for QA, the best setup has been exploited (i.e., $10^4$ reads and an annealing time per read equal to $99\mu$s). The number of runs, for both SA and QA, is 10, and the success rate is given by the ratio between the number of runs in which the expected solution has been discovered and the total number of runs. Instead, the result value represents the ratio between the QUBO image ($x^TQx$) of the solution found and the QUBO image of the expected solution, averaged over the runs (larger values correspond to better performance, since the image value of the expected solution is always negative in these experiments). Regarding the time values, they include only the resolution of the QUBO matrix; specifically, in the case of QA, they correspond to the QPU access time (see \cite{qpu_time} for additional details). Eventually, for QA, the last column (\textit{Average \# exp. sol.}) reports the number of times that the expected solution has been found in a single run, averaged over the runs.

\begin{table}[t]
  \tcaption{Comparison of ES, SA, and QA performances on different problems, using \textit{Exp} datasets of size $N=10^4$, $10^4$ reads for SA, and the best setup for QA ($10^4$ reads, 99 $\mu$s annealing time). The number of runs, for both SA and QA, is 10.}
  \label{tab:comp-performance}
  \centerline{\footnotesize\resizebox{\linewidth}{!}{\begin{tabular}{|c|l|c c | c c c | c c c c|}
    \hline
    \multicolumn{2}{|c|}{} & \multicolumn{2}{c|}{Exhaustive search} & \multicolumn{3}{c|}{Simulated annealing} & \multicolumn{4}{c|}{Quantum annealing} \\
    \hline
    \multirow{2}{*}{n} & \multirow{2}{*}{Problem} & Success & Avg. resol. & Success & Average & Avg. sol. & Success & Average & Avg. resol. & Average \# \\
     & & rate & time (s) & rate & result & time (s) & rate & result & time (s) & exp. sol. \\
    \hline
    3 & MHP       & 1.00 & 0.04   & 1.00 & 1.0000 & 3.40  & 1.00 & 1.0000 & 1.76 & 215.90 \\
    4 & LC4Vars   & 1.00 & 0.42   & 1.00 & 1.0000 & 5.52  & 1.00 & 1.0000 & 1.77 & 11.40  \\
    5 & LC        & 1.00 & 137.53 & 1.00 & 1.0000 & 8.93  & 0.50 & 0.9987 & 1.80 & 0.60   \\
    9 & Waste     & ---  & ---    & 0.00 & 1.0145 & 12.85 & 0.00 & 0.9898 & 2.09 & 0.00   \\
    9 & Waste2P   & ---  & ---    & 0.00 & 0.9999 & 13.15 & 0.00 & 0.9754 & 2.17 & 0.00   \\
    9 & Waste2PDep& ---  & ---    & 0.00 & 0.9998 & 12.39 & 0.00 & 0.9780 & 2.10 & 0.00   \\
    \hline
  \end{tabular}}}
\end{table}

In practice, ES has outperformed both SA and QA on the smallest problems, i.e., MHP and LC4Vars, always finding the minimum in less time. However, due to its exponential complexity, it has lost the comparison on the LC problem ($n=5$), and has turned out to be too time-consuming on the largest ones.

\begin{table}[b!]
  \tcaption{Comparison of quantum annealing performances on several problems for different values of annealing time, using \textit{Exp} datasets of size $N=10^4$ and $10^4$ reads. The number of runs is 10.}
  \label{tab:comp-annealing-time}
  \centerline{\footnotesize\resizebox{\linewidth}{!}{\begin{tabular}{|c|l| c c c c| c c c c|}
    \hline
    \multicolumn{2}{|c|}{} & \multicolumn{4}{c|}{Annealing time per sample $1\mu$s} & \multicolumn{4}{c|}{Annealing time per sample $99\mu$s} \\
    \hline
    \multirow{2}{*}{n} & \multirow{2}{*}{Problem} & Success & Average & Avg. resol. & Average \# & Success & Average & Avg. resol. & Average \#\\
     & & rate & result & time (s) & exp. sol. & rate & result & time (s) & exp. sol. \\
    \hline
    3 & MHP        & 1.00 & 1.0000 & 0.78 & 304.70 & 1.00 & 1.0000 & 1.76 & 215.90 \\
    4 & LC4Vars    & 0.90 & 0.9997 & 0.81 & 5.20   & 1.00 & 1.0000 & 1.77 & 11.40  \\
    5 & LC         & 0.40 & 0.9980 & 0.87 & 0.50   & 0.50 & 0.9987 & 1.80 & 0.60   \\
    9 & Waste      & 0.00 & 0.9619 & 1.15 & 0.00   & 0.00 & 0.9898 & 2.09 & 0.00   \\
    9 & Waste2P    & 0.00 & 0.9473 & 1.15 & 0.00   & 0.00 & 0.9754 & 2.17 & 0.00   \\
    9 & Waste2PDep & 0.00 & 0.9633 & 1.33 & 0.00   & 0.00 & 0.9780 & 2.10 & 0.00   \\
    \hline
  \end{tabular}}}
\end{table}

Concerning the other methods, QA has always managed to find the optimum solution to the problems with three and four Bayesian variables, also outperforming SA in terms of execution time. Moreover, it has detected the global minimum several times in each run ($10^4$ measurements per run are performed), which suggests that a fewer number of reads could be enough to achieve the same results on these problems. Instead, for the five-variable problem, QA has managed to discover the minimum only half of the times (with the minimum occasionally appearing more than once), whereas SA has always found it at the cost of a slightly higher runtime. Eventually, neither QA nor SA have ever detected the optimum solution to the largest problems ($n=9$). Nevertheless, the quality of the solutions found is good on average, especially of those found by SA, whose execution time has turned out to be slightly higher also in this case. It is also worth highlighting that, for the Waste problem, solutions with a better score than that of the expected solution have been found; this has always happened with SA (the average result value is larger than 1.0) and sometimes with QA. The reason lies in the presence of a node with three parents in the expected solution. Basically, this penalizes the expected solution and makes possible to have other solutions respecting the maximum parent constraint ($m \leq 2$) with a better score.

In addition, the impact of the annealing time on the performance of QA in the same experiments has been analyzed; the results for an annealing time of $1\mu$s and $99\mu$s are reported in \cref{tab:comp-annealing-time}. In practice, a higher annealing time has led to no improvement on the smallest problem (MHP) but has provided better results on average, with only a little additional time required, on all the others. In particular, for the problems with four (LC4Vars) and five (LC) Bayesian variables, it has also provided a higher success rate and a higher number of occurrences of the minimum solution.

Eventually, since SA does not have limits on the number of reads, further experiments have been executed on the Waste2PDep problem with the \textit{Exp} dataset of size $N=10^4$, using a number of reads equal to $10^5$ and $10^6$, respectively. The execution time has increased linearly, but no substantial improvement in the performance has been observed.

\subsection{Divide et impera results}
\label{subsec:divide-et-impera-res}
As mentioned in \cref{subsec:problems}, the divide et impera approach has been tested on two problems used for the evaluation of O'Gorman's algorithm, namely, LC and Waste (only in their main variant), and a new additional problem, i.e., Alarm. The results are presented in the following paragraphs.

\paragraph{Execution speedup and timing}
First of all, the speedup achieved exploiting the technique illustrated in \cref{subsec:exec-speedup} has been analyzed. In particular, for each problem taken into account, one \textit{Exp} dataset of size $N = 10^4$ and one run have been used. Moreover, regarding the number of variables for each subproblem ($k$), all values between three (the minimum reasonable value, see \cref{sec:divide-et-impera}) and $n$ have been tested; it is also worth highlighting that $k = n$ corresponds to the direct application of the implementation of O'Gorman's algorithm. The results are reported in \cref{tab:div-et-imp-exec-speedup}, with the time values including the subproblems formulation and the QUBO matrices construction (thus, neither the subproblems resolution nor the final solution reconstruction). In detail, only LC and Waste have been considered here, since the times without speedup for Alarm would have been unfeasible to collect using the machine available. Concerning LC, the time required has been reduced by $\approx 9$ times on average for the divide et impera approach and $\approx 5.6$ times for O'Gorman's algorithm. Instead, for the Waste problem, the speedup has been more significant due to the higher number of variables and/or subproblems, with an average of $\approx 38.4$ times for the divide et impera approach and $\approx 17$ times for O'Gorman's algorithm.

\begin{table}[ht]
    \tcaption{Speedup achieved for different $k$ values on LC and Waste, using an \textit{Exp} datasets of size $N = 10^4$ and a single run. The time values, expressed in seconds, include the subproblems formulation and the QUBO matrices construction. In particular, D.e.I. = divide et impera, O'G. = O'Gorman.}
    \label{tab:div-et-imp-exec-speedup}
    \centerline{\footnotesize\begin{tabular}{|c|c|c|c|c|c|}\hline
        Problem & $k$ & Subproblems \# & Time (no speedup) & Time (with speedup) & Speedup \\ \hline
        \multirow{3}{1.2cm}{\centering LC\\($n=5$)}    & 3 (D.e.I.) & 10  & 27.66   & 4.54  & 6.09x  \\ 
                               & 4 (D.e.I.) & 5   & 54.4    & 4.55  & 11.96x \\ \cline{2-6}
                               & 5 (O'G.)   & 1   & 21.41   & 3.84  & 5.58x  \\ \hline
        \multirow{7}{1.2cm}{\centering Waste\\($n=9$)} & 3 (D.e.I.) & 84  & 237.5   & 10.27 & 23.13x \\ 
                               & 4 (D.e.I.) & 126 & 1754.35 & 32.62 & 53.78x \\ 
                               & 5 (D.e.I.) & 126 & 2704.54 & 66.02 & 40.97x \\ 
                               & 6 (D.e.I.) & 84  & 2907.48 & 78.8  & 36.90x \\ 
                               & 7 (D.e.I.) & 36  & 2186.56 & 54.91 & 39.82x \\ 
                               & 8 (D.e.I.) & 9   & 858.57  & 23.97 & 35.82x \\ \cline{2-6}
                               & 9 (O'G.)   & 1   & 149.58  & 8.8   & 17x    \\ \hline
    \end{tabular}}
\end{table}

Instead, \cref{tab:div-et-imp-exec-times} reports some statistics computed on analogous time values (including subproblems formulation and QUBO matrices construction), which have been obtained using four different non-\textit{Exp} datasets of size $N = 10^4$. The \textit{Exp} datasets have not been included since they may have a lower number of samples, as explained in \cref{subsec:datasets-gen}. In this case, also the Alarm problem has been considered; indeed, all the time values refer to the optimized code (i.e., with speedup). However, due to the still high times, the maximum $k$ value used for it is 9. Eventually, it is worth highlighting that the limit case corresponding to the direct execution of O'Gorman's algorithm ($k = n$) has not been taken into account here. In detail, the highest average time across $k$ values is determined by both the subproblems size and the number of subproblems. Indeed, the average time per subproblem grows with the subproblem size (see the last column). Moreover, looking at the standard deviation (STD) and the coefficient of variation (CV), it turns out that the variance in the input data does not significantly affect the time values. Specifically, the CV value is always lower than $0.05$.

\begin{table}[ht]
\tcaption{Statistics on subproblems formulation and QUBO matrices construction time for different $k$ values. Specifically, 4 non-\textit{Exp} datasets of size $N = 10^4$ have been used for each problem (one run for each dataset). In addition, the time values, expressed in seconds, refer to the optimized code (i.e., with speedup).}
  \label{tab:div-et-imp-exec-times}
  \centerline{\footnotesize\begin{tabular}{|c|c|c|c|c|c|c|} \hline
    \multirow{2}{*}{Problem} & \multirow{2}{*}{$k$} & \multirow{2}{*}{Subproblems \#} & \multirow{2}{*}{Average time} & \multirow{2}{*}{STD time} & \multirow{2}{*}{CV time} & Average time \\
     & & & & & &  per subproblem \\ \hline
    \multirow{2}{1.2cm}{\centering LC\\($n=5$)} & 3 & 10 & 1.32 & 0.02 & 0.014 & 0.132 \\
     & 4 & 5 & 1.28 & 0.03 & 0.022 & 0.256 \\ \hline
    \multirow{6}{1.2cm}{\centering Waste\\($n=9$)} & 3 & 84 & 3.72 & 0.15 & 0.040 & 0.044 \\
    & 4	& 126 & 8.89 & 0.10 & 0.011 & 0.071 \\
    & 5	& 126 & 16.54 & 0.28 & 0.017 & 0.131 \\
    & 6	& 84 & 18.80 & 0.22 & 0.012 & 0.224 \\
    & 7	& 36 & 13.55 & 0.16 & 0.012 & 0.376 \\
    & 8	& 9 & 5.92 & 0.13 & 0.022 & 0.658 \\ \hline
    \multirow{7}{1.2cm}{\centering Alarm\\($n = 15$)} & 3 & 455 & 26.68 & 0.20 & 0.007 & 0.059 \\
    & 4	& 1365 & 203.15 & 0.62 & 0.003 & 0.149 \\
    & 5	& 3003 & 1063.36 & 4.00 & 0.004 & 0.354 \\
    & 6	& 5005 & 2952.22 & 7.08 & 0.002 & 0.590 \\
    & 7	& 6435 & 6548.12 & 235.37 & 0.036 & 1.018 \\
    & 8	& 6435 & 10361.01 & 489.65 & 0.047 & 1.610 \\
    & 9	& 5005 & 11370.70 & 112.19 & 0.010 & 2.272 \\ \hline
  \end{tabular}}
  \vspace{5pt}
\end{table}

\paragraph{Performance}
To evaluate the performance of the divide et impera approach, the following setup has been used for each problem: one \textit{Exp} dataset of size $N = 10^4$, five runs, and a number of variables for each subproblem ($k$) ranging from three (the minimum reasonable value) to $n$ (the total number of Bayesian variables), with the upper limit representing the direct application of O'Gorman's algorithm. Regarding the methods for solving the QUBO encoding, only SA and QA have been exploited in these experiments. Indeed, ES would have required an unreasonable amount of time to solve the subproblems generated with a high $k$ value. Furthermore, 100 reads and an annealing time equal to 1$\mu$s have been used for QA due to the limited quantum resources available and the high number of subproblems to resolve (considering all experiments). To make a fair comparison, 100 reads have been used also for SA. Eventually, it is worth highlighting that only the second reconstruction strategy developed for the divide et impera approach has been applied in these experiments, as explained in \cref{sec:divide-et-impera}.

\begin{figure}[t!]
  \centering
  \includegraphics[width=0.55\linewidth]{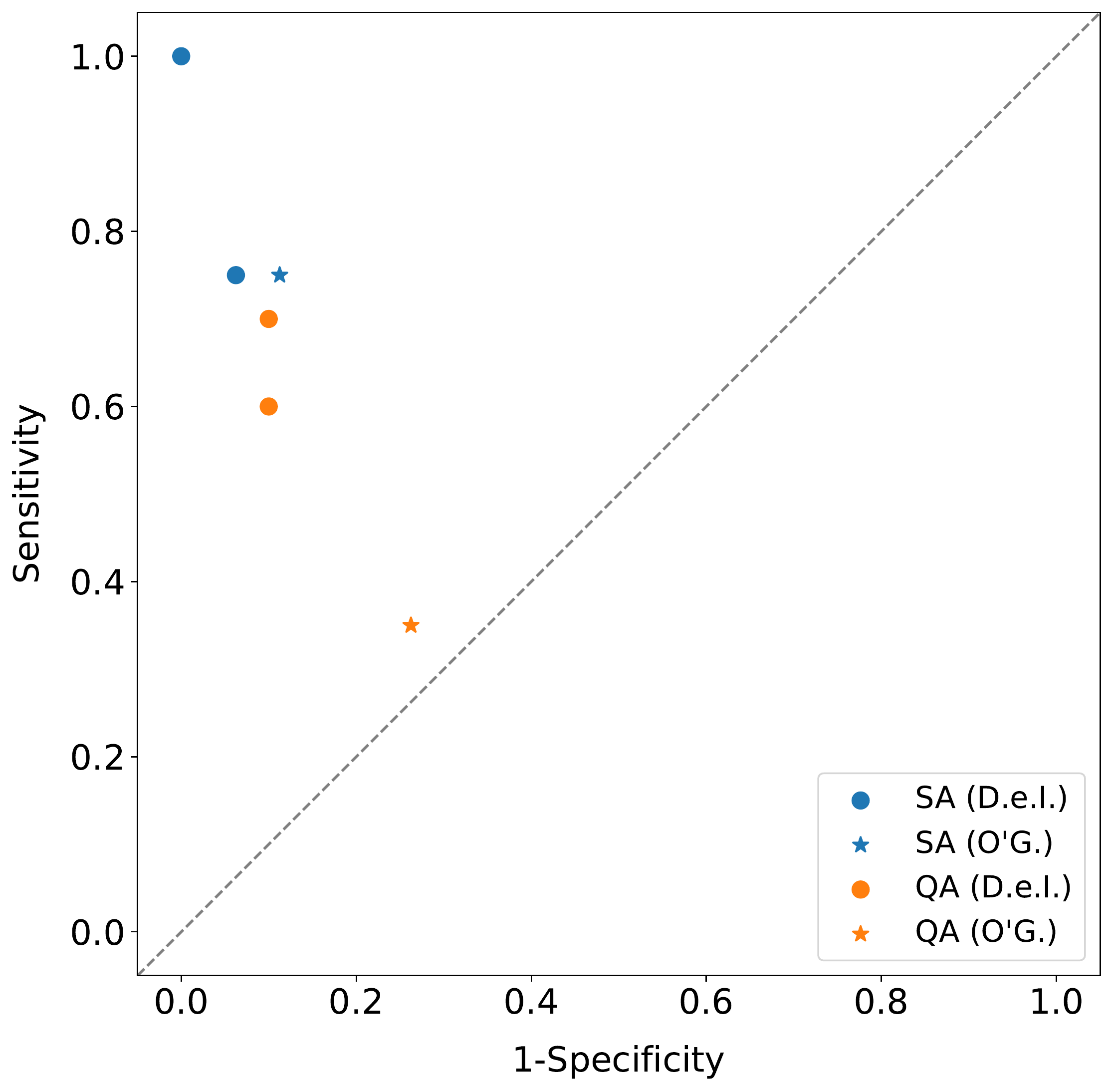}
  \fcaption{Sensitivity versus (1 - Specificity) for the LC problem. This plot results from the data reported in \cref{tab:div-et-imp-lc-res}.}
  \label{fig:lc-roc}
\end{figure}

Starting from LC, the results achieved for it are reported in \cref{tab:div-et-imp-lc-res}. In addition, a "ROC curve"-like plot is provided in \cref{fig:lc-roc}. In practice, SA has turned out to perform better than QA on LC, and the divide et impera approach has outperformed the direct application of O'Gorman's algorithm for both resolution methods. Indeed, the higher the sensitivity and the specificity, the better the result. It is also worth mentioning that SA with $k = 4$ has been able to find the perfect solution (four correct and zero wrong edges) in all five runs. In addition, by looking at the number of unique edges found across all runs (fifth column), it turns out that, for the divide et impera approach, SA tends to find always the same correct and wrong edges, whereas QA shows more variability, as well as O'Gorman's algorithm.

\begin{table}[t]
  \tcaption{Results achieved by the divide et impera approach on the LC problem, for different numbers of variables per subproblem ($k$) and methods (SA/QA), using an \textit{Exp} dataset of size $N = 10^4$, five runs, 100 reads for SA, and 100 reads and 1$\mu$s of annealing time for QA. The last $k$ value (5) corresponds to the direct application of the implementation of O'Gorman's algorithm. In particular, D.e.I. = divide et impera, O'G. = O'Gorman.}
  \label{tab:div-et-imp-lc-res}
  \centerline{\footnotesize\resizebox{\linewidth}{!}{\begin{tabular}{|c|c|c|c c c c c|c|c|c|c|} \hline
    \multicolumn{12}{|c|}{\multirow{2}{*}{LC ($n=5$, edges\,=\,4)}} \\
    \multicolumn{12}{|c|}{} \\ \hline
    $k$ & Method & Metric & \multicolumn{5}{|c|}{\# for each run} & \# unique & Average \# & Sensitivity & Specificity \\ \hline \hline
      & \multirow{2}{*}{SA} & Correct edges & 2 & 2 & 4 & 4 & 3 & 4 & 3 & \multirow{2}{*}{0.75} & \multirow{2}{*}{0.94} \\
    3 & & Wrong edges & 2 & 2 & 0 & 0 & 1 & 2 & 1 &  &  \\ \cline{2-12}
    (D.e.I.) & \multirow{2}{*}{QA} & Correct edges & 1 & 4 & 4 & 1 & 2 & 4 & 2.4 & \multirow{2}{*}{0.60} & \multirow{2}{*}{0.90} \\
      & & Wrong edges & 3 & 0 & 0 & 3 & 2 & 4 & 1.6 &  &  \\ \hline \hline
      & \multirow{2}{*}{SA} & Correct edges & 4 & 4 & 4 & 4 & 4 & 4 & 4 & \multirow{2}{*}{1.00} & \multirow{2}{*}{1.00} \\
    4 & & Wrong edges & 0 & 0 & 0 & 0 & 0 & 0 & 0 & & \\ \cline{2-12}
    (D.e.I.) & \multirow{2}{*}{QA} & Correct edges & 3 & 3 & 3 & 3 & 2 & 4 & 2.8 & \multirow{2}{*}{0.70} & \multirow{2}{*}{0.90} \\
      & & Wrong edges & 2 & 2 & 1 & 1 & 2 & 5 & 1.6 &  &  \\ \hline \hline
      & \multirow{2}{*}{SA} & Correct edges & 3 & 4 & 3 & 3 & 2 & 4 & 3 & \multirow{2}{*}{0.75} & \multirow{2}{*}{0.89} \\
    5 & & Wrong edges & 2 & 0 & 2 & 2 & 3 & 4 & 1.8 &  &  \\ \cline{2-12}
    (O'G.) & \multirow{2}{*}{QA} & Correct edges & 1 & 2 & 0 & 2 & 2 & 3 & 1.4 & \multirow{2}{*}{0.35} & \multirow{2}{*}{0.74} \\
      & & Wrong edges & 5 & 5 & 6 & 2 & 3 & 11 & 4.2 &  &  \\
    \hline
  \end{tabular}}}
\end{table}

\begin{figure}[t!]
  \centering
  \vspace{15pt}
  \includegraphics[width=0.55\linewidth]{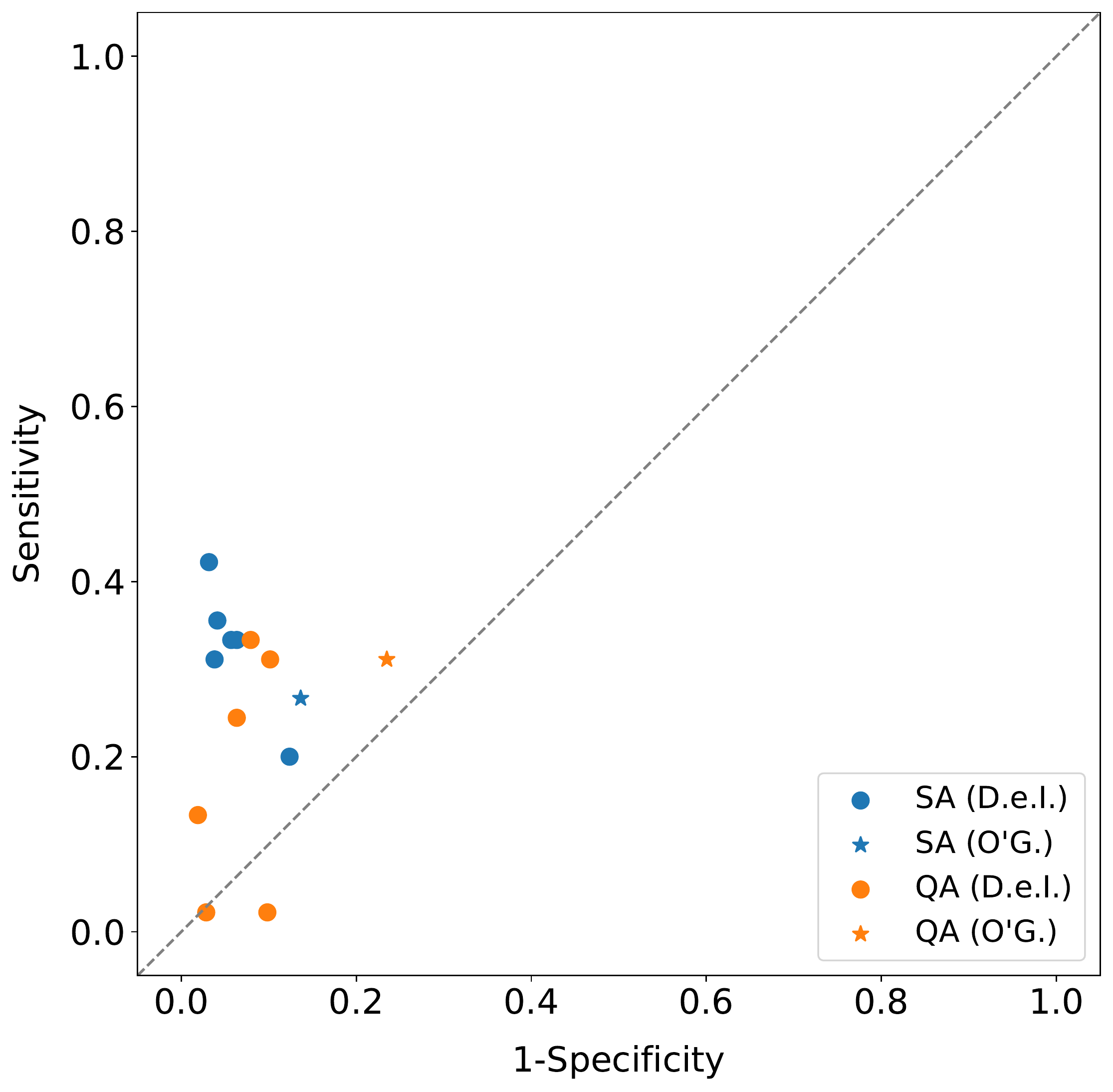}
  \fcaption{Sensitivity versus (1 - Specificity) for the Waste problem. This plot results from the data reported in \cref{tab:div-et-imp-waste-res}.}
  \label{fig:waste-roc}

  \vspace{25pt}

  \centering
  \includegraphics[width=0.55\linewidth]{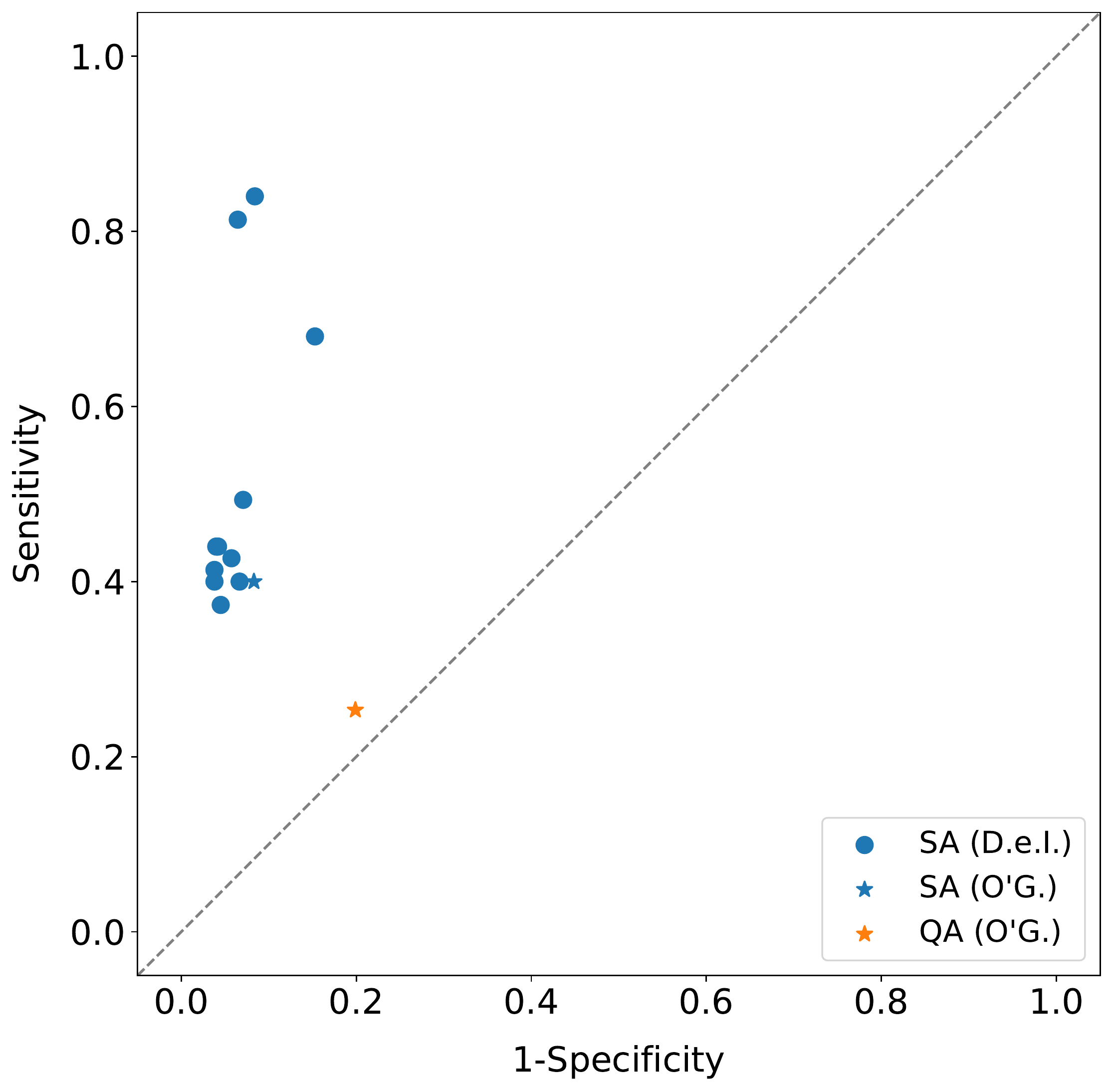}
  \fcaption{Sensitivity versus (1 - Specificity) for the Alarm problem. This plot results from the data reported in \cref{tab:div-et-imp-alarm-res}.}
  \label{fig:alarm-roc}
  \vspace{15pt}
\end{figure}

Concerning the Waste problem, whose results are reported in \cref{tab:div-et-imp-waste-res} and displayed in \cref{fig:waste-roc}, the overall performance is worse for both SA and QA. Specifically, also in this case, SA has performed better than QA overall. Only for $k=3$, QA has been able to achieve better results on average. Instead, the superiority of the divide et impera approach w.r.t. the direct application of O'Gorman's algorithm has turned out to be less marked. In detail, for SA, the divide et impera approach has won the comparison for almost all (but not all) $k$ values. Regarding QA, O'Gorman's algorithm has achieved a relatively high sensitivity (w.r.t. all QA results), but the corresponding specificity is quite low. In the end, for both methods (SA and QA), it is possible to find a $k$ value for which the divide et impera approach has performed better than O'Gorman's algorithm. Actually, for this problem, the perfect solution has never been found. The best results have been achieved by SA with $k = 6$, which has discovered four correct edges out of nine in almost all runs and only two wrong edges on average. In addition, the maximum number of correct edges that have been found in a single run is equal to 6 (SA with $k = 4$). However, it is worth remarking that the problem in question has been subjected to a discretization procedure and includes a node with three parents. Eventually, in general, the observations made for LC on the number of unique edges found turn out to be valid also for the Waste problem. The only difference lies in the correct edges found by the divide et impera approach with QA; indeed, they tend to be the same across runs.

\begin{table}[t!]
  \tcaption{Results achieved by the divide et impera approach on the Waste problem, for different numbers of variables per subproblem ($k$) and methods (SA/QA), using an \textit{Exp} dataset of size $N = 10^4$, five runs, 100 reads for SA, and 100 reads and 1$\mu$s of annealing time for QA. The last $k$ value (9) corresponds to the direct application of the implementation of O'Gorman's algorithm. In particular, D.e.I. = divide et impera, O'G. = O'Gorman, Sens. = sensitivity, Spec. = specificity.}
  \label{tab:div-et-imp-waste-res}
  \centerline{\footnotesize\resizebox{\linewidth}{!}{\begin{tabular}{|c|c|c|c c c c c|c|c|c|c|} \hline
    \multicolumn{12}{|c|}{\multirow{2}{*}{Waste ($n=9$, edges\,=\,9)}} \\
    \multicolumn{12}{|c|}{} \\ \hline
    $k$ & Method & Metric & \multicolumn{5}{|c|}{\# for each run} & \# unique & Average \# & Sens. & Spec. \\ \hline \hline
      & \multirow{2}{*}{SA} & Correct edges & 2 & 2 & 2 & 2 & 1 & 2 & 1.8 & \multirow{2}{*}{0.20} & \multirow{2}{*}{0.88} \\
    3 & & Wrong edges & 8 & 7 & 7 & 8 & 9 & 10 & 7.8 &  &  \\ \cline{2-12}
    (D.e.I.) & \multirow{2}{*}{QA} & Correct edges & 1 & 4 & 3 & 4 & 2 & 6 & 2.8 & \multirow{2}{*}{0.31} & \multirow{2}{*}{0.90} \\
      & & Wrong edges & 8 & 6 & 6 & 5 & 7 & 13 & 6.4 &  &  \\ \hline \hline
      & \multirow{2}{*}{SA} & Correct edges & 2 & 6 & 2 & 3 & 2 & 6 & 3 & \multirow{2}{*}{0.33} & \multirow{2}{*}{0.94} \\
    4 & & Wrong edges & 5 & 1 & 5 & 4 & 5 & 5 & 4 &  &  \\ \cline{2-12}
    (D.e.I.) & \multirow{2}{*}{QA} & Correct edges & 4 & 1 & 3 & 3 & 4 & 6 & 3 & \multirow{2}{*}{0.33} & \multirow{2}{*}{0.92} \\
      & & Wrong edges & 4 & 7 & 4 & 5 & 5 & 10 & 5 &  &  \\ \hline \hline
      & \multirow{2}{*}{SA} & Correct edges & 2 & 3 & 4 & 3 & 3 & 5 & 3 & \multirow{2}{*}{0.33} & \multirow{2}{*}{0.94} \\
    5 & & Wrong edges & 5 & 4 & 3 & 2 & 4 & 7 & 3.6 &  &  \\ \cline{2-12}
    (D.e.I.) & \multirow{2}{*}{QA} & Correct edges & 3 & 5 & 2 & 0 & 1 & 5 & 2.2 & \multirow{2}{*}{0.24} & \multirow{2}{*}{0.94} \\
      & & Wrong edges & 4 & 2 & 5 & 4 & 5 & 10 & 4.0 &  &  \\ \hline \hline
      & \multirow{2}{*}{SA} & Correct edges & 4 & 4 & 4 & 4 & 3 & 5 & 3.8 & \multirow{2}{*}{0.42} & \multirow{2}{*}{0.97} \\
    6 & & Wrong edges & 2 & 1 & 3 & 1 & 3 & 4 & 2 &  &  \\ \cline{2-12}
    (D.e.I.) & \multirow{2}{*}{QA} & Correct edges & 1 & 1 & 1 & 2 & 1 & 3 & 1.2 & \multirow{2}{*}{0.13} & \multirow{2}{*}{0.98} \\
      & & Wrong edges & 1 & 2 & 1 & 0 & 2 & 5 & 1.2 &  &  \\ \hline \hline
      & \multirow{2}{*}{SA} & Correct edges & 5 & 4 & 1 & 3 & 3 & 5 & 3.2 & \multirow{2}{*}{0.36} & \multirow{2}{*}{0.96} \\
    7 & & Wrong edges & 1 & 1 & 5 & 3 & 3 & 6 & 2.6 &  &  \\ \cline{2-12}
    (D.e.I.) & \multirow{2}{*}{QA} & Correct edges & 0 & 0 & 0 & 1 & 0 & 1 & 0.2 & \multirow{2}{*}{0.02} & \multirow{2}{*}{0.97} \\
      & & Wrong edges & 2 & 3 & 1 & 2 & 1 & 8 & 1.8 &  &  \\ \hline \hline
      & \multirow{2}{*}{SA} & Correct edges & 2 & 1 & 4 & 5 & 2 & 5 & 2.8 & \multirow{2}{*}{0.31} & \multirow{2}{*}{0.96} \\
    8 & & Wrong edges & 3 & 4 & 2 & 0 & 3 & 6 & 2.4 &  &  \\ \cline{2-12}
    (D.e.I.) & \multirow{2}{*}{QA} & Correct edges & 1 & 0 & 0 & 0 & 0 & 1 & 0.2 & \multirow{2}{*}{0.02} & \multirow{2}{*}{0.90} \\
      & & Wrong edges & 4 & 8 & 6 & 6 & 7 & 25 & 6.2 &  &  \\ \hline \hline
      & \multirow{2}{*}{SA} & Correct edges & 3 & 3 & 3 & 1 & 2 & 5 & 2.4 & \multirow{2}{*}{0.27} & \multirow{2}{*}{0.86} \\
    9 & & Wrong edges & 9 & 8 & 8 & 9 & 9 & 25 & 8.6 &  &  \\ \cline{2-12}
    (O'G.) & \multirow{2}{*}{QA} & Correct edges & 2 & 2 & 3 & 3 & 4 & 7 & 2.8 & \multirow{2}{*}{0.31} & \multirow{2}{*}{0.77} \\
      & & Wrong edges & 15 & 16 & 15 & 12 & 16 & 49 & 14.8 &  &  \\ \hline
  \end{tabular}}}
\end{table}

Finally, the results related to the Alarm problem are reported in \cref{tab:div-et-imp-alarm-res} and shown in \cref{fig:alarm-roc}. In particular, the divide et impera approach with QA has not been evaluated in this case because the number of subproblems is really high and the sequential submission of numerous QUBO problems to the D-Wave's annealer tends to fail due to connectivity issues (from D-Wave's side), invalidating the run. As for the other problems, the divide et impera approach has outperformed the direct application of O'Gorman's algorithm. Indeed, the latter has won the comparison (with a worse specificity) only for $k = 12$. In addition, SA has achieved far better results than O'Gorman's algorithm with QA. Concerning the quality of the solution found, the best results have been achieved by SA with $k = 4$, which has been capable of detecting $12.6$ correct edges out of 15 on average (note that the problem includes a variable with four parents). However, the number of wrong edges is quite high ($16.4$ on average). The same configuration has also discovered the highest number of correct edges (13). Eventually, it is worth making two last observations: the number of edges (both correct and wrong) detected by the divide et impera approach tends to decrease by increasing the value of $k$; the divide et impera approach tends to discover always the same correct and wrong edges across runs (look at the fifth column), whereas O'Gorman's algorithm exhibits more variability. Actually, the wrong edges for $k=3$ and the correct edges for O'Gorman's algorithm with SA represent two exceptions.

\begin{table}[t!]
  \tcaption{Results achieved by the divide et impera approach on the Alarm problem, for different numbers of variables per subproblem ($k$), using an \textit{Exp} dataset of size $N = 10^4$, five runs, 100 reads for SA, and 100 reads and 1$\mu$s of annealing time for QA. The last $k$ value (15) corresponds to the direct application of the implementation of O'Gorman's algorithm. In particular, D.e.I. = divide et impera, O'G. = O'Gorman, Sens. = sensitivity, Spec. = specificity.}
  \label{tab:div-et-imp-alarm-res}
  \centerline{\footnotesize\resizebox{\linewidth}{!}{\begin{tabular}{|c|c|c|c c c c c|c|c|c|c|} \hline
    \multicolumn{12}{|c|}{\multirow{2}{*}{Alarm ($n=15$, edges\,=\,15)}} \\
    \multicolumn{12}{|c|}{} \\ \hline
    $k$ & Method & Metric & \multicolumn{5}{|c|}{\# for each run} & \# unique & Average \# & Sens. & Spec. \\ \hline \hline
    3 & \multirow{2}{*}{SA} & Correct edges & 10 & 10 & 11 & 10 & 10 & 12 & 10.2 & \multirow{2}{*}{0.68} & \multirow{2}{*}{0.85} \\
    (D.e.I.) & & Wrong edges & 31 & 31 & 30 & 29 & 28 & 38 & 29.8 &  &  \\\hline \hline
    4 & \multirow{2}{*}{SA} & Correct edges & 12 & 13 & 13 & 12 & 13 & 13 & 12.6 & \multirow{2}{*}{0.84} & \multirow{2}{*}{0.92} \\
    (D.e.I.) & & Wrong edges & 17 & 16 & 17 & 16 & 16 & 20 & 16.4 &  &  \\ \hline \hline
    5 & \multirow{2}{*}{SA} & Correct edges & 12 & 12 & 12 & 12 & 13 & 13 & 12.2 & \multirow{2}{*}{0.81} & \multirow{2}{*}{0.94} \\
    (D.e.I.) & & Wrong edges & 13 & 13 & 13 & 12 & 12 & 15 & 12.6 &  &  \\ \hline \hline
    6 & \multirow{2}{*}{SA} & Correct edges & 8 & 7 & 7 & 7 & 8 & 8 & 7.4 & \multirow{2}{*}{0.49} & \multirow{2}{*}{0.93} \\
    (D.e.I.) & & Wrong edges & 14 & 14 & 14 & 14 & 13 & 14 & 13.8 &  &  \\ \hline \hline
    7 & \multirow{2}{*}{SA} & Correct edges & 6 & 6 & 6 & 6 & 6 & 6 & 6 & \multirow{2}{*}{0.40} & \multirow{2}{*}{0.93} \\
    (D.e.I.) & & Wrong edges & 13 & 13 & 13 & 13 & 13 & 13 & 13 &  &  \\\hline \hline
    8 & \multirow{2}{*}{SA} & Correct edges & 7 & 6 & 6 & 7 & 6 & 7 & 6.4 & \multirow{2}{*}{0.43} & \multirow{2}{*}{0.94} \\
    (D.e.I.) & & Wrong edges & 11 & 11 & 11 & 11 & 12 & 12 & 11.2 &  &  \\ \hline \hline
    9 & \multirow{2}{*}{SA} & Correct edges & 7 & 6 & 7 & 7 & 6 & 7 & 6.6 & \multirow{2}{*}{0.44} & \multirow{2}{*}{0.96} \\
    (D.e.I.) & & Wrong edges & 8 & 9 & 7 & 7 & 9 & 10 & 8.0 &  &  \\ \hline \hline
    10 & \multirow{2}{*}{SA} & Correct edges & 6 & 6 & 7 & 7 & 7 & 7 & 6.6 & \multirow{2}{*}{0.44} & \multirow{2}{*}{0.96} \\
    (D.e.I.) & & Wrong edges & 9 & 8 & 8 & 7 & 7 & 9 & 7.8 &  &  \\ \hline \hline
    11 & \multirow{2}{*}{SA} & Correct edges & 6 & 7 & 7 & 7 & 6 & 7 & 6.6 & \multirow{2}{*}{0.44} & \multirow{2}{*}{0.96} \\
    (D.e.I.) & & Wrong edges & 8 & 8 & 8 & 8 & 9 & 9 & 8.2 &  &  \\ \hline \hline
    12 & \multirow{2}{*}{SA} & Correct edges & 5 & 6 & 6 & 5 & 6 & 6 & 5.6 & \multirow{2}{*}{0.37} & \multirow{2}{*}{0.95} \\
    (D.e.I.) & & Wrong edges & 9 & 9 & 8 & 9 & 9 & 10 & 8.8 &  &  \\ \hline \hline
    13 & \multirow{2}{*}{SA} & Correct edges & 5 & 7 & 6 & 7 & 6 & 8 & 6.2 & \multirow{2}{*}{0.41} & \multirow{2}{*}{0.96} \\
    (D.e.I.) & & Wrong edges & 9 & 6 & 8 & 7 & 7 & 9 & 7.4 &  &  \\ \hline \hline
    14 & \multirow{2}{*}{SA} & Correct edges & 5 & 7 & 6 & 7 & 5 & 9 & 6 & \multirow{2}{*}{0.40} & \multirow{2}{*}{0.96} \\
    (D.e.I.) & & Wrong edges & 10 & 8 & 6 & 6 & 7 & 11 & 7.4 &  &  \\ \hline \hline
      & \multirow{2}{*}{SA} & Correct edges & 4 & 5 & 6 & 6 & 9 & 9 & 6 & \multirow{2}{*}{0.40} & \multirow{2}{*}{0.92} \\
    15 & & Wrong edges & 18 & 16 & 16 & 16 & 15 & 55 & 16.2 &  &  \\ \cline{2-12}
    (O'G.) & \multirow{2}{*}{QA} & Correct edges & 2 & 4 & 5 & 7 & 1 & 12 & 3.8 & \multirow{2}{*}{0.25} & \multirow{2}{*}{0.80} \\
      & & Wrong edges & 39 & 41 & 36 & 40 & 38 & 128 & 38.8 &  &  \\ \hline
  \end{tabular}}}
\end{table}

\newpage

\section{Conclusion}
\label{sec:conclusion}
\noindent
In this work, we have presented an implementation in Python of the algorithm proposed by O'Gorman et al. for solving the BNSL problem on a quantum annealer, a divide et impera approach that allows addressing BNSL instances with a higher number of variables, a complexity analysis of them, and their experimental evaluation. In detail, to make O'Gorman's formulation effectively usable, algebraic manipulations have been applied to the computation of the local scores $s_i(\varPi_i(B_s))$. Moreover, a simplified lower bound has been introduced for the penalty value $\delta_{consist}$, and the best setup of the $\alpha_{ijk}$ hyperparameters has been empirically determined. The results achieved in the experiments have demonstrated that O'Gorman's algorithm can be effectively used to reconstruct Bayesian networks of small sizes ($n <= 5$) with less than three parents per node ($m < 3$). Instead, in presence of more Bayesian variables ($n = 9$), the algorithm performance have turned out to be worse. Indeed, good quality solutions (in terms of QUBO image value) have been found, but not the correct one. It is also worth remarking that one of these larger problems includes a node with three parents. In addition, the linear dependency between the dataset size and the QUBO matrix construction time has been confirmed. Eventually, QA (using the best annealing parameters) has been able to achieve comparable or slightly worse results with respect to SA, proving the competitiveness of the current annealing architectures on this task.

Concerning the divide et impera approach, which has been developed to overcome the limitation on the problem size dictated by the available annealing devices, the results have demonstrated that it performs better than the direct application of O'Gorman's algorithm. Indeed, in all problems considered, for all resolution methods tested, there is more than one $k$ value for which the divide et impera approach has achieved better results; actually, in almost all cases, these $k$ values represent the majority. Instead, in general, the quality (in terms of resulting Bayesian network) of the solutions found has turned out to be not optimal. However, non-ideal annealing parameters have been used for QA due to the limited quantum resources at our disposal, and the number of reads for SA has also been reduced (w.r.t. the value used for the experiments on O'Gorman's algorithm) for a fair comparison. Moreover, in this second set of experiments, unlike in the first one, SA has performed definitely better than QA; nevertheless, this is probably related to the less-performing parameters used here. Finally, the experiments on the subproblems formulation and QUBO matrices construction time have confirmed the effectiveness of the technique used to speed up the execution (and also the independence of the times from the variance in the input data).

Future work includes the following possibilities: testing the divide et impera approach on Bayesian problems whose size is larger than the maximum size embeddable in the Pegasus architecture using O'Gorman's algorithm; evaluating the aforementioned approach with QA using more-performing annealing parameters; analysing the impact of considering only part of the subproblems of size $k$. We conclude by reminding that the code of both the implementation of O'Gorman's algorithm and the divide et impera approach is available under the GPLv2 licence \cite{ogorman_implementation_github,divide_et_impera_implementation_github}.

\section*{Acknowledgements}
\noindent
This work was supported by Q@TN, the joint lab between University of Trento, FBK-Fondazione Bruno Kessler, INFN-National Institute for Nuclear Physics and CNR-National Research Council. In addition, the authors gratefully acknowledge the J\"ulich Supercomputing Center (\url{https://www.fz-juelich.de/ias/jsc}) for funding this project by providing computing time through the J\"ulich UNified Infrastructure of Quantum computing (JUNIQ) on the D-Wave quantum annealer.

\end{document}

%% file: symbols_and_commands.tex
\bmdefine{\bz}{z}
\bmdefine{\bTheta}{\Theta}

\def\beq{\begin{eqnarray}}
\def\eeq{\end{eqnarray}}

\let\oldparagraph\paragraph
\renewcommand{\paragraph}[1]{\oldparagraph{#1}\mbox{}\\}